\documentclass[preprint]{aastex63}

\pdfoutput=1 
\usepackage{appendix}
\usepackage{amsmath,amstext}
\usepackage[figure,figure*]{hypcap}

\def\wotan{{\tt W\={o}tan}}

\providecommand{\rsun}{\ensuremath{\,R_{\odot}}}

\providecommand{\mj}{\ensuremath{\,M_{\rm J}}}
\providecommand{\rj}{\ensuremath{\,R_{\rm J}}}
\providecommand{\me}{\ensuremath{\,M_{\rm \oplus}}}
\providecommand{\re}{\ensuremath{\,R_{\rm \oplus}}}

\newcommand{\tess}{\emph{TESS}}

\usepackage{scrextend}
\let\orgautoref\autoref
\renewcommand{\autoref}
        {\def\equationautorefname{Eq.}%
         \def\figureautorefname{Fig.}%
         \def\sectionautorefname{Sect.}%
         \def\subsectionautorefname{Sect.}%
         \def\subsubsectionautorefname{Sect.}%
         \orgautoref}

\usepackage{changepage} 
\usepackage{rotating}

\newcommand{\mos}{\,m\,s$^{-1}$}
\newcommand{\kms}{\,km\,s$^{-1}$}

\submitjournal{Astronomical Journal}

\shorttitle{TOI-1842b}
\shortauthors{Wittenmyer et al.}

\graphicspath{{./}{figures/}}

\begin{document}

\title{TOI-1842b: A Transiting Warm Saturn Undergoing Re-Inflation around an Evolving Subgiant}

\correspondingauthor{Robert A. Wittenmyer}
\email{rob.w@usq.edu.au}

\author[0000-0001-9957-9304]{Robert A. Wittenmyer}
\affiliation{University of Southern Queensland, Centre for Astrophysics, USQ Toowoomba, West Street, QLD 4350 Australia}


\author[0000-0003-3964-4658]{Jake T. Clark}
\affiliation{University of Southern Queensland, Centre for Astrophysics, USQ Toowoomba, West Street, QLD 4350 Australia}

\author[0000-0002-0236-775X]{Trifon Trifonov}
\affiliation{Max-Planck-Institut fur Astronomie, Konigstuhl 17, 69117 Heidelberg, Germany}

\author[0000-0003-3216-0626]{Brett C. Addison}
\affiliation{University of Southern Queensland, Centre for Astrophysics, USQ Toowoomba, West Street, QLD 4350 Australia}

\author{Duncan J. Wright}
\affiliation{University of Southern Queensland, Centre for Astrophysics, USQ Toowoomba, West Street, QLD 4350 Australia}

\author[0000-0002-3481-9052]{Keivan G. Stassun}
\affiliation{Vanderbilt University, Department of Physics \& Astronomy, Nashville, TN 37235, USA}

\author[0000-0002-1160-7970]{Jonathan Horner}
\affiliation{University of Southern Queensland, Centre for Astrophysics, USQ Toowoomba, West Street, QLD 4350 Australia}

\author[0000-0001-6508-5736]{Nataliea Lowson}
\affiliation{University of Southern Queensland, Centre for Astrophysics, USQ Toowoomba, West Street, QLD 4350 Australia}

\author[0000-0003-0497-2651]{John Kielkopf}
\affil{Department of Physics and Astronomy, University of Louisville, Louisville, KY 40292, USA}

\author[0000-0002-7084-0529]{Stephen R. Kane}
\affiliation{Department of Earth and Planetary Sciences, University of California, Riverside, CA 92521, USA}

\author[0000-0002-8864-1667]{Peter Plavchan}
\affiliation{Department of Physics \& Astronomy, George Mason University, 4400 University Drive MS 3F3, Fairfax, VA 22030, USA}

\author[0000-0002-1836-3120]{Avi Shporer}
\affil{Department of Physics and Kavli Institute for Astrophysics and Space Research, Massachusetts Institute of Technology, Cambridge, MA 02139, USA}


\author[0000-0003-3491-6394]{Hui Zhang}
\affil{School of Astronomy and Space Science, Key Laboratory of Modern Astronomy and Astrophysics in Ministry of Education, Nanjing University, Nanjing 210046, Jiangsu, China}


\author[0000-0003-2649-2288]{Brendan P. Bowler}
\affiliation{Department of Astronomy, The University of Texas at Austin, Austin, TX 78712, USA}

\author{Matthew W. Mengel}
\affil{University of Southern Queensland, Centre for Astrophysics, West Street, Toowoomba, QLD 4350 Australia}

\author[0000-0002-4876-8540]{Jack Okumura}
\affil{University of Southern Queensland, Centre for Astrophysics, West Street, Toowoomba, QLD 4350 Australia}

\author[0000-0003-2935-7196]{Markus Rabus}
\affiliation{Departamento de Matem\'atica y F\'isica Aplicadas, Universidad Cat\'olica de la Sant\'isima Concepci\'on, Alonso de Rivera 2850, Concepci\'on, Chile}
\affiliation{Las Cumbres Observatory, 6740 Cortona Dr., Ste. 102, Goleta, CA 93117, USA}
\affiliation{Department of Physics, University of California, Santa Barbara, CA 93106-9530, USA}

\author[0000-0002-5099-8185]{Marshall C. Johnson}
\affiliation{Las Cumbres Observatory, 6740 Cortona Dr., Ste. 102, Goleta, CA 93117, USA}

\author[0000-0002-8590-007X]{Daniel Harbeck}
\affiliation{Las Cumbres Observatory, 6740 Cortona Dr., Ste. 102, Goleta, CA 93117, USA}


\author[0000-0003-1001-0707]{Ren\'{e} Tronsgaard}
\affiliation{DTU Space, National Space Institute, Technical University of Denmark, Elektrovej 328, DK-2800 Kgs. Lyngby, Denmark}
 
\author[0000-0003-1605-5666]{Lars A. Buchhave}
\affiliation{DTU Space, National Space Institute, Technical University of Denmark, Elektrovej 328, DK-2800 Kgs. Lyngby, Denmark}


\author[0000-0001-6588-9574]{Karen A.\ Collins}
\affiliation{Center for Astrophysics \textbar \ Harvard \& Smithsonian, 60 Garden Street, Cambridge, MA 02138, USA}

\author[0000-0003-2781-3207]{Kevin I.\ Collins}
\affiliation{Department of Physics \& Astronomy, George Mason University, 4400 University Drive MS 3F3, Fairfax, VA 22030, USA}

\author[0000-0002-4503-9705]{Tianjun Gan}
\affiliation{Department of Astronomy and Tsinghua Centre for Astrophysics, Tsinghua University, Beijing 100084, China}"

\author[0000-0002-4625-7333]{Eric L.\ N.\ Jensen}
\affiliation{Department of Physics \& Astronomy, Swarthmore College, Swarthmore PA 19081, USA}

\author[0000-0002-2532-2853]{Steve~B.~Howell}
\affil{NASA Ames Research Center, Moffett Field, CA 94035, USA}

\author[0000-0001-9800-6248]{E. Furlan}
\affiliation{NASA Exoplanet Science Institute, Caltech/IPAC, Mail Code 100-22, 1200 E. California Blvd.,
Pasadena, CA 91125, USA}

\author[0000-0003-2519-6161]{Crystal~L.~Gnilka}
\affil{NASA Ames Research Center, Moffett Field, CA 94035, USA}

\author[0000-0002-9903-9911]{Kathryn V. Lester}
\affil{NASA Ames Research Center, Moffett Field, CA 94035, USA}

\author[0000-0001-7233-7508]{Rachel A. Matson}
\affiliation{U.S. Naval Observatory, 3450 Massachusetts Avenue NW, Washington, D.C. 20392, USA}

\author[0000-0003-1038-9702]{Nicholas~J.~Scott}
\affil{NASA Ames Research Center, Moffett Field, CA 94035, USA}



\author[0000-0003-2058-6662]{George~R.~Ricker}
\affiliation{Department of Physics and Kavli Institute for Astrophysics and Space Research, Massachusetts Institute of Technology, Cambridge, MA 02139, USA}

\author[0000-0001-6763-6562]{Roland~Vanderspek}
\affiliation{Department of Physics and Kavli Institute for Astrophysics and Space Research, Massachusetts Institute of Technology, Cambridge, MA 02139, USA}

\author[0000-0001-9911-7388]{David~W.~Latham}
\affiliation{Center for Astrophysics \textbar \ Harvard \& Smithsonian, 60 Garden St, Cambridge, MA 02138, USA}

\author[0000-0002-6892-6948]{S.~Seager}
\affiliation{Department of Physics and Kavli Institute for Astrophysics and Space Research, Massachusetts Institute of Technology, Cambridge, MA 02139, USA}
\affiliation{Department of Earth, Atmospheric and Planetary Sciences, Massachusetts Institute of Technology, Cambridge, MA 02139, USA}
\affiliation{Department of Aeronautics and Astronautics, MIT, 77 Massachusetts Avenue, Cambridge, MA 02139, USA}

\author[0000-0002-4265-047X]{Joshua~N.~Winn}
\affiliation{Department of Astrophysical Sciences, Princeton University, Princeton, NJ 08540, USA}

\author[0000-0002-4715-9460]{Jon~M.~Jenkins}
\affiliation{NASA Ames Research Center, Moffett Field, CA 94035, USA}

\author{Alexander~Rudat}
\affiliation{Department of Physics and Kavli Institute for Astrophysics and Space Research, Massachusetts Institute of Technology, Cambridge, MA 02139, USA}

\author[0000-0003-1309-2904]{Elisa~V.~Quintana}
\affiliation{NASA Goddard Space Flight Center, 8800 Greenbelt Road, Greenbelt, MD 20771, USA}

\author{David~R.~Rodriguez}
\affiliation{Space Telescope Science Institute, 3700 San Martin Drive, Baltimore, MD, 21218, USA}

\author[0000-0003-1963-9616]{Douglas~A.~Caldwell}
\affiliation{NASA Ames Research Center, Moffett Field, CA 94035, USA}
\affiliation{SETI Institute, Mountain View, CA 94043, USA}

\author[0000-0002-8964-8377]{Samuel~N.~Quinn}
\affiliation{Center for Astrophysics \textbar \ Harvard \& Smithsonian, 60 Garden Street, Cambridge, MA 02138, USA}

\author[0000-0002-2482-0180]{Zahra~Essack}
\affiliation{Department of Earth, Atmospheric and Planetary Sciences, Massachusetts Institute of Technology, Cambridge, MA 02139, USA}
\affiliation{Kavli Institute for Astrophysics and Space Research, Massachusetts Institute of Technology, Cambridge, MA 02139, USA}

\author[0000-0002-0514-5538]{Luke~G.~Bouma}
\affiliation{Department of Astrophysical Sciences, Princeton University, 4 Ivy Lane, Princeton, NJ 08544, USA}

\begin{abstract}

The imminent launch of space telescopes designed to probe the atmospheres of exoplanets has prompted new efforts to prioritise the thousands of transiting planet candidates for follow-up characterisation.  We report the detection and confirmation of TOI-1842b, a warm Saturn identified by \textit{TESS} and confirmed with ground-based observations from \textsc{Minerva}-Australis, NRES, and the Las Cumbres Observatory Global Telescope.  This planet has a radius of $1.04^{+0.06}_{-0.05}$\rj, a mass of $0.214^{+0.040}_{-0.038}$\mj, an orbital period of $9.5739^{+0.0002}_{-0.0001}$ days, and an extremely low density ($\rho$=0.252$\pm$0.091 g cm$^{-3}$).  TOI-1842b has among the best known combinations of large atmospheric scale height (893 km) and host-star brightness ($J=8.747$ mag), making it an attractive target for atmospheric characterisation.  As the host star is beginning to evolve off the main sequence, TOI-1842b presents an excellent opportunity to test models of gas giant re-inflation.  The primary transit duration of only 4.3 hours also makes TOI-1842b an easily-schedulable target for further ground-based atmospheric characterisation.  

\end{abstract}


\section{Introduction} \label{sec:intro}


The tireless efforts of exoplanetary astronomers have, for more than three decades, relentlessly banished the darkness shrouding the mysteries of planetary system formation and evolution in the Solar neighbourhood.  The stars that our ancestors once thought of as distant campfires are now known to host retinues of worlds in a riotous array of sizes, compositions, and orbital properties.  Of late, we have fuelled this revolution with deceptively small space-borne robotic sentries; the \textit{Kepler} \citep{borucki10} and now \textit{TESS} (Transiting Exoplanet Survey Satellite, Ricker et al. 2015) spacecraft have delivered 2981 new confirmed exoplanets from over 6000 candidates\footnote{Planet data from the NASA Exoplanet Archive at \url{https://exoplanetarchive.ipac.caltech.edu/}, accessed 2021 August 25}.  

The Level 1 mission requirement of \textit{TESS} is to obtain mass and radius measurements for 50 planets smaller than 4\re\ \citep{TESSRick}, and at the end of its two-year prime mission in 2020 July, nearly 40 such Level 1 planets have been confirmed \citep[e.g.][]{TESS1,TESS2,TESS3,TESS4,Trifonov2021}.  As is common in astronomy, where the most attention is lavished on discoveries at the margins of detectability, transits of larger planets are often overshadowed by their smaller kin.  However, such warm giants from the \textit{TESS} mission are also scientifically valuable \citep[e.g.][]{warm1,warm2,warm3,Schlecker2020}.  These giant planets, with orbital periods between about 5 and 10 days, reside far enough from their host star that tidal circularisation has not yet erased any orbital eccentricity or inclination imposed by post-formation migration mechanisms \citep{dawson18}.  Such planets are thus important laboratories for understanding the means by which close-in giant planets are delivered to their present locations.  

In a similar vein, planets orbiting stars of different types than the Sun remain under-represented.  This arose first from the technical requirements of the radial velocity method, which favoured Solar-type stars (late F through K dwarfs) due to the Doppler information contained within their abundant and narrow spectral lines \citep[e.g.][]{51Peg,RV1,RV2,2004A&A...423..385P,RV4}.  As the transit technique flowered in the age of \textit{Kepler}, the bias toward Solar-type stars was reinforced to give the mission the best chance of achieving its goal of detecting Earth-size planets.  Again, the hotter and more evolved stars were spurned for their younger and cooler counterparts.  Evolved stars in particular are heavily disfavoured in transit searches since their larger stellar radii diminish the transit signals of planets.  The advent of purpose-built infrared high-precision spectrographs \citep{HPF,carmenes,IRD,veloce} has fuelled an outsized interest in the coolest stars, stoked heavily by the profligacy with which the universe produces systems of multiple small planets around these M dwarfs \citep[e.g.][]{cool1,wolf1061,cool3,cool4}, and because habitable-zone planets are more accessible.  

However, the hotter, higher-mass stars (A type through mid-F dwarfs) have yet to receive much attention from the exoplanet community.  While on the main sequence, these stars offer a limited number of useful spectral lines for radial velocity measurement.  Some progress toward understanding the exoplanet population hosted by these stars has been achieved by studying their evolved counterparts, since as subgiants and low-luminosity giants, the stars cool down and slow their rotation, resulting in a forest of narrow spectral lines that permit precise velocity measurements \citep[e.g.][]{BigJohnson,bowler10,jones11,sato13,ppps8}.  A conundrum has been the apparent lack of giant planets in orbits shorter than $\sim$100 days \citep{johnson10,reffert15,k239,jones21}.  Contrary to popular belief, the hot- and warm-Jupiter populations have not been engulfed by gastronomically rapacious host stars -- the low-luminosity giants targeted by the major surveys have typical radii of less than 10\rsun, so close-in giant planets would remain intact until the last $\sim$1\% of their lifetimes \citep{vl09,kunitomo11}.  Transit surveys have been successful in confirming the presence of close-in giant planets around main-sequence A and F stars \citep[e.g.][]{snellen09,mascara1,kelt24,toi778}, and preliminary estimates of the occurrence rates suggest that there is no significant dependence on host-star mass \citep{zhou19}.



In this paper, we report the discovery and confirmation of TOI-1842b, a rare short-period giant planet orbiting an evolved star.  Section 2 details the various photometric and spectroscopic observations, and in Section 3, we describe the properties of the host star.  Section 4 gives the results of the joint fitting, and we conclude in Section 5.

\section{Observations and Data Reduction}\label{sec:observations}

In this section, we describe the \textit{TESS} photometry, follow-up ground-based photometry, and spectroscopic observations used to establish the planetary nature of TOI-1842b.  

\subsection{Photometric Observations} 

\subsubsection{\textit{TESS} Light Curve} 
\label{tesslc}

TOI-1842 (TIC\,404505029, \citealt{TIC19}) was observed by \textit{TESS} on Camera 1 in Sector 23 in 2-minute cadence mode nearly continuously between 2020 March 19 and 2020 April 15. The \textit{TESS} Science Processing Operations Center (SPOC) pipeline \citep[see,][for a description of the SPOC pipeline]{jenkins2016} was used to process the photometric data, resulting in two versions of the light curves: Simple Aperture Photometry \citep[SAP, see,][]{Twicken2010,2020ksci.rept....6M} and Presearch Data Conditioning \citep[PDC, see,][]{2012PASP..124..985S,2014PASP..126..100S,2012PASP..124.1000S}. Both versions of the light curves were downloaded from the NASA's Mikulski Archive for Space Telescopes (MAST) for analysis.  The SPOC conducted a transit search of the PDC light curve with an adaptive, noise-compensating matched filter \citep{jenkins2002, jenkins2010}, triggering on two transits of TOI-1842b. No additional transiting planet signatures were identified in a search of the residual light curve.

The transiting planet candidate was released as a candidate \textit{TESS} Object of Interest (TOI) and designated TOI-1842.01 by the \textit{TESS} Science Office based on model fit results and a passing grade on all diagnostic tests in the SPOC Data Validation (DV) report \citep{2018PASP..130f4502T,2019PASP..131b4506L} for Sector 23.  These diagnostic tests included a difference image centroiding analysis that located the source of the transit signature to within 1.6$\pm$2.5 arcsec of TOI-1842.  In total, two transits were observed by \textit{TESS} with a duration of $\sim$4.4 hours, a depth of $\sim$3150 parts per million (ppm), and a period of $\sim9.57$\,days, with the first detected transit occurring on BJD$_{\mathrm{TDB}}$ 2458933. Joint analysis (see Section~\ref{sec:Results}) of the radial velocity data, \textit{TESS} light curve, and ground-based follow-up light curve from the Las Cumbres Observatory Global Telescope (LCOGT) confirm that the planet has a period of $\sim9.57$\,days (half the original period published in the SPOC DV report of $\sim19.15$\,days), resulting in one transit being missed by \textit{TESS} during the data downlink gap as shown in Figure~\ref{Transit_plots}.

We removed all quality-flagged data from the SAP and PDC light curves before detrending and fitting them. The light curves were then split into a total of four segments, with each segment split at the spacecraft momentum dumps\footnote{As provided in the data release notes found at \url{https://archive.stsci.edu/tess/tess_drn.html}} that occurred 8.0\,days after the beginning of each orbit, to minimize the offset effects on the light curves during the fitting analysis. The transits were then masked and a $5\sigma$ median filter was applied to the out-of-transit data to remove remaining outliers in each light curve segment. After masking the transits and applying a median filter, we detrended each light curve segment using an rspline with iterative sigma-clipping and a filter window set to 4.0 times the planet period using the Python package \wotan{} \citep{wotan}. Lastly, the light curve segments were normalized with the mean of the out-of-transit flux before performing the fitting analysis as described in Section~\ref{sec:Results}.  In the joint fit, we choose the detrended PDC light curve over the detrended SAP light curve as it has already been corrected for instrumental and systematic effects as well as for light dilution by the SPOC pipeline.  


\subsubsection{Las Cumbres Observatory}

We obtained three transits of TOI-1842.01 in Pan-STARSS $z$-short band with an exposure time of 30 s on UTC 2020 May 30, 2021 February 13, and 2021 April 2 from Las Cumbres Observatory Global Telescope (LCOGT) \citep{Brown13} 1.0\,m network nodes. The first photometric observation was taken at the South Africa Astronomical Observatory while the latter two were done at the Cerro Tololo Interamerican Observatory. We used the {\tt  TESS Transit Finder}, which is a customized version of the {\tt Tapir} software package \citep{Jensen:2013}, to schedule our transit observations. The first transit was scheduled using a period of 9.5740475 days (half the original period), and on an epoch that corresponds to a possibly missed transit in the \textit{TESS} data gap (see Section \ref{tesslc}). The $4096\times4096$ LCOGT SINISTRO cameras have an image scale of $0\farcs389$ per pixel, resulting in a $26\arcmin\times26\arcmin$ field of view. The images were calibrated by the standard LCOGT {\tt BANZAI} pipeline \citep{McCully:2018}, and photometric data were extracted using {\tt AstroImageJ} \citep{Collins:2017}. The images were defocused and have typical stellar point-spread-functions with a full-width-half-maximum (FWHM) of $\sim 5\farcs 7$, and circular apertures with radius $\sim 6\farcs 6$ were used to extract the differential photometry. A transit with depth $\sim 3000$ ppm was detected on target, showing that the true period is half the originally published $\sim19.1$ day period (see Section \ref{sec:Results}). The latter two follow up observations confirmed the return of the signal on target and at the 9.57d period.

\subsection{Spectroscopic Observations}

TOI-1842 has been observed by three facilities for stellar parameter determination and precise radial velocity follow up.  Here we give details about the observations from each instrument.  All radial velocities used in this analysis are given in Table~\ref{tab:allRV}.

\subsubsection{MINERVA-Australis}

We carried out the spectroscopic observations of TOI-1842 using the {\textsc{Minerva}}-Australis facility \citep{2018arXiv180609282W,addison2019,TOI257}.  {\sc {\textsc{Minerva}}}-Australis consists of an array of four independently operated 0.7\,m CDK700 telescopes situated at the Mount Kent Observatory in Queensland, Australia \citep{addison2019}.  Each telescope simultaneously feeds stellar light via fiber optic cables to a single KiwiSpec R4-100 high-resolution ($R=80,000$) spectrograph \citep{2012SPIE.8446E..88B} with wavelength coverage from 480 to 620\,nm.  Wavelength calibration is achieved with a simultaneous Th-Ar arc lamp observation through a calibration fibre.  We obtained a total of 119 individual spectra from 2020 May 12 to 2020 Aug 3, with exposure times of 45-60 minutes, distributed amongst the three then-operational {\textsc{Minerva}}-Australis telescopes as follows: T1 -- 31; T4 -- 42; T5 -- 46.  Radial velocities (Table~\ref{tab:allRV}) were derived for each telescope by cross-correlation, where the template being matched is the mean spectrum of each telescope.  The radial velocity data suggested that the true orbital period is one-half of the period that had been initially reported by the \textit{TESS} team. This finding was contemporaneously verified by TFOP SG1 photometry as described above. 

\subsubsection{LCOGT/NRES}

We obtained 12 spectra of TOI-1842 using the Network of Robotic Echelle Spectrographs \citep[NRES;][]{Siverd18} on the Las Cumbres Observatory \citep[LCOGT;][]{Brown13} telescope network.  NRES is a set of four identical echelle spectograph units at observatory sites around the world.  NRES achieves a resolving power of $R\sim53,000$ with continuous coverage over 3900-8600 \AA.  We reduced the data and measured radial velocities using the \texttt{CERES} pipeline \citep{Brahm17}.  We obtained the data between 2020 June 7 and July 22 UT.  Five of the spectra were obtained with the NRES unit at the Southern African Astronomical Observatory, South Africa, and seven with the unit at Wise Observatory, Israel.

\subsubsection{FIES}

A single spectrum of TOI-1842 was obtained on the night of 2020 May 27 (UT) using the FIbre-fed \'{E}chelle Spectrograph \citep[FIES;][]{Telting14} at the Nordic Optical Telescope \citep[NOT;][]{djupvik10}.  We observed with the high-resolution fiber, which offers a resolution of $R \sim 67,000$ and has continuous wavelength coverage from about 3760 to 8220\,\AA.  The data were reduced following \citet{Buchhave10}.

\startlongtable
\begin{deluxetable}{cccc}
\tabletypesize{\scriptsize}
\tablecaption{Radial Velocities for TOI-1842}
\label{tab:allRV}
\tablehead{
\colhead{Time} & \colhead{Velocity} & \colhead{Uncertainty} & \colhead{Fibre/Telescope}\\
\colhead{[BJD]} & \colhead{[\mos]} & \colhead{[\mos]} & \colhead{}}
\startdata
  \multicolumn4c{MINERVA-Australis} \\
   2458981.99578  &    3730.8  &   17.4  & 4 \\
   2458982.03844  &    3654.9  &   15.5  & 4 \\
   2458995.05865  &    3677.3  &   12.6  & 4 \\
   2458995.05865  &    3695.3  &   10.1  & 5 \\
   2458995.10137  &    3667.2  &   13.5  & 4 \\
   2458995.10137  &    3670.7  &   11.8  & 5 \\
   2458996.01626  &    3669.3  &   12.6  & 4 \\
   2458996.01626  &    3700.1  &    9.3  & 5 \\
   2458996.05897  &    3616.9  &   17.1  & 4 \\
   2458996.05897  &    3667.9  &   13.9  & 5 \\
   2458997.97378  &    3703.6  &   12.3  & 5 \\
   2458997.97378  &    3725.2  &   15.1  & 4 \\
   2458998.01650  &    3709.2  &   11.8  & 4 \\
   2458998.01650  &    3713.9  &    9.1  & 5 \\
   2458998.87519  &    3629.3  &   11.9  & 1 \\
   2458998.87519  &    3669.8  &   10.6  & 4 \\
   2458998.87519  &    3742.2  &   10.9  & 5 \\
   2458998.91789  &    3646.5  &   10.0  & 1 \\
   2458998.91789  &    3687.4  &   11.2  & 4 \\
   2458998.91789  &    3694.1  &    9.1  & 5 \\
   2458999.87477  &    3623.5  &   12.1  & 1 \\
   2458999.87477  &    3678.9  &   11.2  & 4 \\
   2458999.87477  &    3690.6  &   10.4  & 5 \\
   2458999.91749  &    3647.2  &   10.7  & 1 \\
   2458999.91749  &    3675.0  &   10.8  & 4 \\
   2458999.91749  &    3704.5  &    9.1  & 5 \\
   2459000.87519  &    3664.1  &   13.8  & 5 \\
   2459000.87519  &    3667.9  &   15.0  & 4 \\
   2459000.91791  &    3683.9  &   10.6  & 5 \\
   2459000.91791  &    3689.3  &   13.2  & 4 \\
   2459001.87494  &    3658.3  &   19.0  & 4 \\
   2459001.87494  &    3691.3  &   15.4  & 5 \\
   2459001.91766  &    3642.7  &   13.5  & 5 \\
   2459001.91766  &    3695.1  &   21.4  & 4 \\
   2459002.87503  &    3672.1  &    9.7  & 5 \\
   2459002.87503  &    3680.8  &   14.1  & 4 \\
   2459002.95885  &    3614.2  &    8.9  & 1 \\
   2459002.95885  &    3665.9  &    9.4  & 5 \\
   2459002.95885  &    3693.0  &   16.6  & 4 \\
   2459003.00157  &    3585.6  &    8.6  & 1 \\
   2459003.00157  &    3628.5  &   13.4  & 4 \\
   2459003.00157  &    3663.3  &    9.5  & 5 \\
   2459003.96260  &    3602.3  &    8.8  & 1 \\
   2459003.96260  &    3667.6  &   13.2  & 4 \\
   2459003.96260  &    3672.4  &    9.2  & 5 \\
   2459004.00531  &    3596.1  &    8.8  & 1 \\
   2459004.00531  &    3669.9  &   13.3  & 4 \\
   2459004.00531  &    3683.0  &    9.5  & 5 \\
   2459004.04903  &    3623.1  &    9.3  & 1 \\
   2459004.04903  &    3651.2  &   13.7  & 4 \\
   2459004.04903  &    3695.8  &    9.5  & 5 \\
   2459005.88421  &    3629.4  &    9.5  & 1 \\
   2459005.88421  &    3702.1  &    8.9  & 5 \\
   2459005.92693  &    3639.4  &    9.1  & 1 \\
   2459005.92693  &    3705.7  &    8.6  & 5 \\
   2459006.87339  &    3649.4  &   15.2  & 1 \\
   2459006.87339  &    3703.0  &   11.9  & 5 \\
   2459006.87339  &    3716.6  &   21.2  & 4 \\
   2459006.91611  &    3614.4  &   43.2  & 4 \\
   2459006.91611  &    3628.9  &   16.5  & 1 \\
   2459006.91611  &    3717.5  &   11.4  & 5 \\
   2459008.87830  &    3699.9  &   14.7  & 1 \\
   2459008.87830  &    3713.8  &    8.8  & 5 \\
   2459008.92102  &    3646.0  &   10.2  & 1 \\
   2459008.92102  &    3681.5  &    8.3  & 5 \\
   2459015.87987  &    3680.3  &    9.7  & 5 \\
   2459017.98362  &    3637.8  &    9.1  & 1 \\
   2459017.98362  &    3692.8  &    9.2  & 5 \\
   2459017.98362  &    3698.2  &   12.2  & 4 \\
   2459022.89711  &    3658.2  &   11.1  & 1 \\
   2459022.89711  &    3665.3  &   12.6  & 4 \\
   2459022.89711  &    3679.3  &    9.4  & 5 \\
   2459022.93984  &    3589.5  &    9.2  & 1 \\
   2459022.93984  &    3679.9  &    9.2  & 5 \\
   2459022.93984  &    3684.7  &   12.2  & 4 \\
   2459023.93098  &    3619.7  &   13.1  & 4 \\
   2459023.93098  &    3673.0  &    9.2  & 5 \\
   2459024.95263  &    3619.9  &    8.2  & 1 \\
   2459024.95263  &    3667.9  &   12.5  & 4 \\
   2459024.95263  &    3697.6  &    9.1  & 5 \\
   2459024.99541  &    3631.9  &    8.7  & 1 \\
   2459024.99541  &    3678.1  &   11.9  & 4 \\
   2459024.99541  &    3715.8  &    9.1  & 5 \\
   2459025.90669  &    3688.4  &   12.8  & 4 \\
   2459025.90669  &    3689.3  &    8.9  & 5 \\
   2459025.94941  &    3691.3  &   13.3  & 4 \\
   2459025.94941  &    3701.4  &    9.0  & 5 \\
   2459026.91155  &    3622.5  &    8.4  & 1 \\
   2459026.91155  &    3653.9  &   15.3  & 4 \\
   2459026.91155  &    3694.2  &    8.9  & 5 \\
   2459026.95427  &    3644.1  &    8.4  & 1 \\
   2459026.95427  &    3664.0  &   13.2  & 4 \\
   2459026.95427  &    3720.5  &    8.8  & 5 \\
   2459027.88275  &    3644.1  &    9.3  & 1 \\
   2459027.88275  &    3674.3  &   12.7  & 4 \\
   2459027.88275  &    3711.1  &    9.1  & 5 \\
   2459027.92549  &    3639.9  &    8.7  & 1 \\
   2459027.92549  &    3705.3  &    9.0  & 5 \\
   2459027.92549  &    3708.8  &   12.7  & 4 \\
   2459059.90505  &    3650.7  &   33.1  & 1 \\
   2459059.90505  &    3661.4  &   17.2  & 5 \\
   2459059.90505  &    3705.6  &   15.1  & 4 \\
   2459059.93701  &    3608.4  &   19.3  & 1 \\
   2459059.93701  &    3673.9  &   15.8  & 4 \\
   2459059.93701  &    3730.6  &   22.3  & 5 \\
   2459061.88858  &    3635.0  &   18.9  & 1 \\
   2459061.88858  &    3692.4  &   67.1  & 4 \\
   2459061.88858  &    3716.7  &   16.3  & 5 \\
   2459061.92055  &    3671.7  &   51.9  & 1 \\
   2459061.92055  &    3677.5  &   20.8  & 5 \\
   2459062.88413  &    3659.1  &   58.8  & 1 \\
   2459062.88413  &    3683.6  &   15.9  & 4 \\
   2459062.88413  &    3715.9  &   23.8  & 5 \\
   2459062.91608  &    3696.2  &   17.9  & 1 \\
   2459062.91608  &    3731.7  &   16.0  & 4 \\
   2459062.91608  &    3796.8  &   32.4  & 5 \\
   2459064.89018  &    3676.3  &   26.6  & 1 \\
   2459064.89018  &    3736.8  &   15.2  & 4 \\
   2459064.89018  &    3742.1  &   17.3  & 5 \\
  \hline
  \multicolumn4c{NRES} \\
  2459021.21141522  &  3398.2  &  17.7 &  SAAO \\  
  2459036.23065811  &  3352.7  &  13.1 &  SAAO \\
  2459047.21646529  &  3392.2  &  14.2 &  SAAO \\  
  2459046.25208904  &  3322.8  &  13.1 &  SAAO \\  
  2459053.22874907  &  3412.1  &  14.0 &  SAAO \\   
  2459008.29685002  &  3492.6  &  26.1 &  Wise \\  
  2459013.35257704  &  3287.8  &  36.3 &  Wise \\ 
  2459023.27975045  &  3288.4  &  18.4 &  Wise \\ 
  2459025.32730386  &  3244.6  &  28.4 &  Wise \\  
  2459038.28150024  &  3212.0  &  17.9 &  Wise \\  
  2459035.27860615  &  3207.5  &  22.6  &  Wise \\ 
  2459041.2807924   &  3109.9  &  36.6  &  Wise \\  
\enddata
\end{deluxetable}

\section{Stellar Properties of TOI-1842}\label{sec:thestar}

\subsection{Spectroscopic Stellar Parameter Determination}

We derived stellar parameters for TOI-1842 using spectra from \textsc{Minerva}-Australis as fully detailed in \citet{TOI257}.  The spectroscopic parameters $T_{\rm eff}$, $\log g$, and overall metallicity [M/H] were derived using \texttt{iSpec} \citep{ispec2014,ispec2019}.  Our derived $T_{\rm eff}$, $\log g$, and [M/H] values were then used as input for the Bayesian isochrone modeler \texttt{isochrones} \citep{morton15,montet15} that uses the Dartmouth Stellar Evolution Database \citep{dotter08}. The \texttt{isochrones} modelling also included \textit{Gaia} $G, G_{BP}, G_{RP}$ magnitudes; {\it 2MASS} $J, H$, and $K_S$ magnitudes along with \textit{Gaia} EDR3 parallax values and their respective errors.

The extracted FIES spectrum was compared to a library of synthetic templates to measure effective temperature $T_{\rm eff}$, surface gravity $\log g$, projected rotational velocity $v \sin i$, and metallicity [M/H] (a solar mix of metals, rather than Fe only).  Our analysis follows \citet{Buchhave12} and \citet{Buchhave14}, using five spectral orders spanning a wavelength range from 5065 to 5320 \AA.  

The spectroscopic and derived physical parameters for TOI-1842 are given in Table~\ref{tab:star}.  In summary, we find that TOI-1842 is a slightly evolved subgiant star which is somewhat more massive and metal-rich than the Sun.  While the stellar mass from the TIC is $2\sigma$ different from that derived with the Minerva-Australis spectrum, we note that the TIC mass does not use a spectroscopic log $g$, which can lead to discrepancies in derived masses \citep[e.g.][]{clark21}.  The \texttt{isochrones}-derived stellar mass and radius are adopted for the joint modelling as presented in the next section. 

%
%
\begin{deluxetable}{lccc}
\tablewidth{0.45\textwidth}
\tablecolumns{4}
\tablecaption{Stellar parameters for TOI-1842. \label{tab:star}}
\tablehead{
\colhead{Parameter}  & \colhead{Value}  & \colhead{Reference} \\
}
\startdata
Right Ascension (h:m:s)   & 13:27:51.04 & 1 \\ \hline
Declination (d:m:s)       & +09:01:50.32 & 1 \\ \hline
Distance (pc)             & 223.47~$\pm$~2.24 & 1 \\ \hline
$V$ (mag)                 & 9.81~$\pm$~0.03 & 2 \\ \hline 
$T$ (mag)                 & 9.2673~$\pm$~0.0061  & 3 \\ \hline
$T_{\rm eff}$ (K)         & 6115~$\pm$~119 & 3 \\
                          & 6239~$\pm$~50 & 4 \\
                          & \textbf{6230~$\pm$~50} & 5 \\ \hline
$\log g$ (cm\,s$^{-2}$)   & 3.876~$\pm$~0.080 & 3 \\
                          & 4.25$\pm$0.10 & 4 \\ 
                          & \textbf{4.10~$\pm$~0.10} & 5 \\ \hline
$R_{\star}$ ($R_{\odot}$) & 2.05~$\pm$~0.09 & 3 \\
                          & \textbf{2.02~$\pm$~0.05} & 5 \\
                          & 2.013~$\pm$~0.045 & 6 \\ \hline
$L_{\star}$ ($L_{\odot}$) & 5.29~$\pm$~0.22  & 3 \\
                          & \textbf{5.50~$\pm$~0.31} & 5 \\ \hline
$M_{\star}$ ($M_{\odot}$) & 1.15~$\pm$~0.15 & 3 \\
                          & \textbf{1.46~$\pm$~0.03} & 5 \\
                          & 1.78~$\pm$~0.42 & 6 \\ \hline
Metallicity, [M/H]        & 0.39~$\pm$~0.08 & 4 \\
                          & 0.22~$\pm$~0.05 & 5 \\ \hline
Age (Gyr)                 & 2.49~$\pm$~0.22 & 5 \\ \hline
$\rho_{\star}$ (g cm$^{-3}$) & 0.189~$\pm$~0.041 & 3 \\
                          & 0.237~$\pm$~0.021 & 6 \\
                          & 0.25~$\pm$~0.02 & 5 \\ \hline
$v \sin i$ (\kms)         & 4.23~$\pm$~0.20        & 5 \\ 
                          & 7.6~$\pm$~0.5 & 4 \\ \hline
\enddata
\raggedright
\tablerefs{1. \cite{GAIADR2}; 2. \cite{2000Hog}; 3. \cite{TIC19}, 4. FIES spectrum (this work), 5. \textsc{Minerva}-Australis spectrum (this work), 6. SED fit (Section~\ref{sec:sed}). }
\tablecomments{Adopted values are shown in bold.}

\end{deluxetable}


\subsection{Spectral Energy Distribution}\label{sec:sed}

\begin{figure}[!ht]
    \centering
    \includegraphics[width=\linewidth,trim=90 75 80 90,clip]{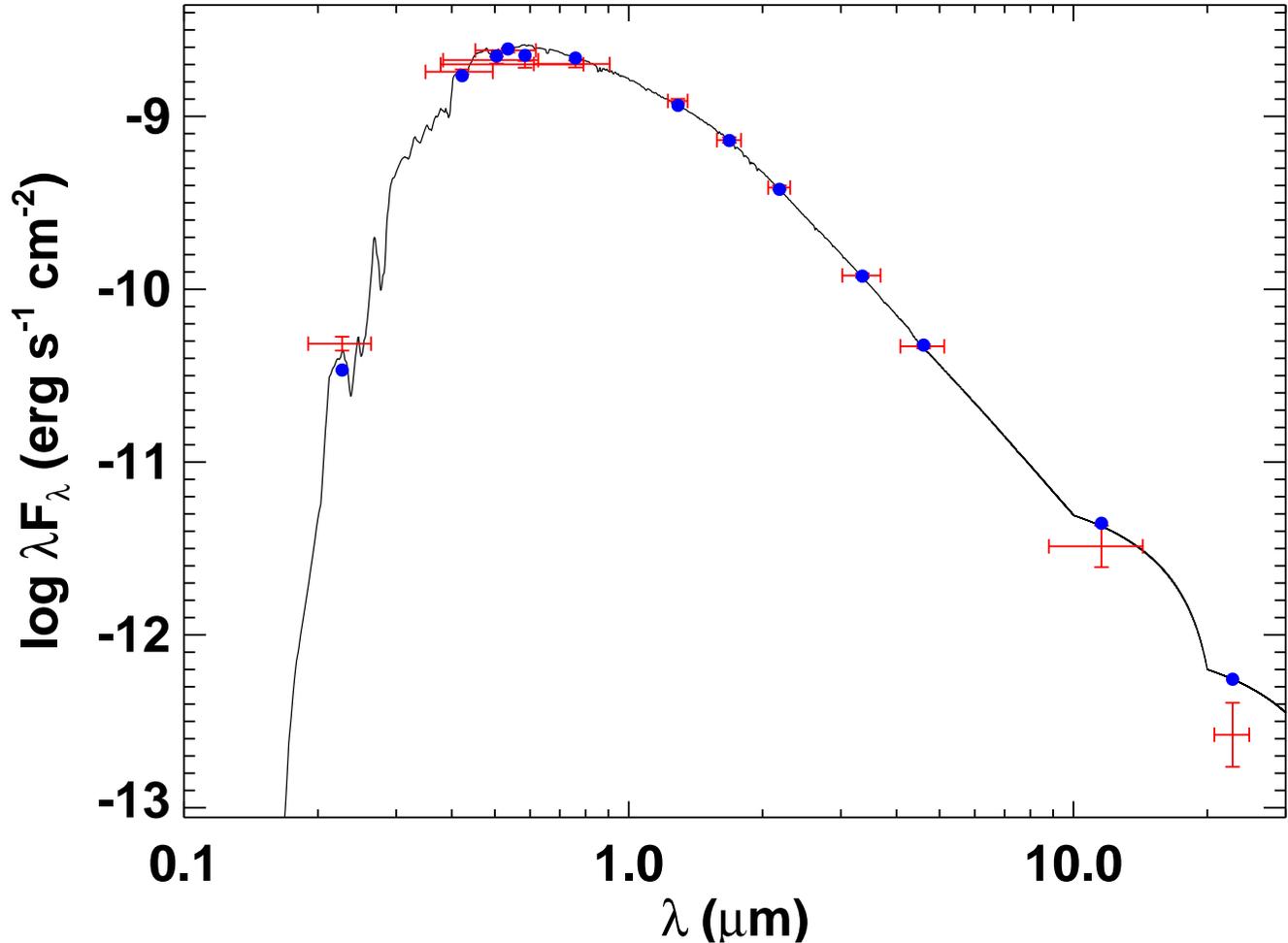}
\caption{Spectral energy distribution of TOI~1842. Red symbols represent the observed photometric measurements, where the horizontal bars represent the effective width of the passband. Blue symbols are the model fluxes from the best-fit Kurucz atmosphere model (black).  \label{fig:sed}}
\end{figure}

In order to provide an initial empirical constraint on the stellar radius and mass, we performed an analysis of the broadband spectral energy distribution (SED) together with the {\it Gaia\/} DR2 parallax following the procedures described in \citet{Stassun:2016,Stassun:2017,Stassun:2018}. We used the NUV flux from {\it GALEX}, the $B_T V_T$ magnitudes from {\it Tycho-2}, the $JHK_S$ magnitudes from {\it 2MASS}, the W1--W4 magnitudes from {\it WISE}, and the $G G_{BP} G_{RP}$ magnitudes from {\it Gaia}. Together, the available photometry spans the full stellar SED over the wavelength range 0.2--22~$\mu$m (see Figure~\ref{fig:sed}).

We performed a fit using Kurucz stellar atmosphere models, adopting the effective temperature ($T_{\rm eff}$), surface gravity ($\log g$), and metallicity ([Fe/H]) from the spectroscopically determined values. The extinction ($A_V$) was allowed to vary but limited to the maximum line-of-sight extinction from the dust maps of \citet{Schlegel:1998}. The resulting fit is very good (Figure~\ref{fig:sed}) with a reduced $\chi^2$ of 1.4. Integrating the (unreddened) model SED gives the bolometric flux at Earth of $F_{\rm bol} =  3.278 \pm 0.076 \times 10^{-9}$ erg~s$^{-1}$~cm$^{-2}$. Taking the $F_{\rm bol}$ and $T_{\rm eff}$ together with the {\it Gaia\/} DR2 parallax, adjusted by $+0.08$~mas to account for the systematic offset reported by \citet{StassunTorres:2018}, gives the stellar radius as $R_\star = 2.013 \pm 0.045$~R$_\odot$. 

In addition, we can compute an empirical estimate for the stellar mass via the empirical stellar radius above together with the spectroscopic $\log g$, which gives $M_\star = 1.8 \pm 0.4$~M$_\odot$, consistent with the empirical eclipsing-binary based relations of \citet{Torres:2010}, which give $M = 1.37 \pm 0.08$~M$_\odot$. The mass and radius together give a mean stellar density of $\rho_\star = 0.24 \pm 0.02$~g~cm$^{-3}$.  These estimates are also consistent with the mass and radius derived from the \textsc{Minerva}-Australis spectra.

Finally, from the spectroscopic $v\sin i$ and the stellar radius we obtain an upper limit on the stellar rotation period, $P_{\rm rot}/\sin i = 13.4 \pm 0.9$~d. Using the empirical gyrochronology relations of \citet{Mamajek:2008}, this then provides upper limits on the stellar age of $\approx$1.9~Gyr ($1\sigma$) and $\approx$1.5~Gyr ($3\sigma$).  The lack of lithium absorption in the FIES spectrum also places a lower age limit of $\sim$800 Myr.  



\subsection{High Resolution Imaging} 

\begin{figure}
  \includegraphics[width=8.0cm]{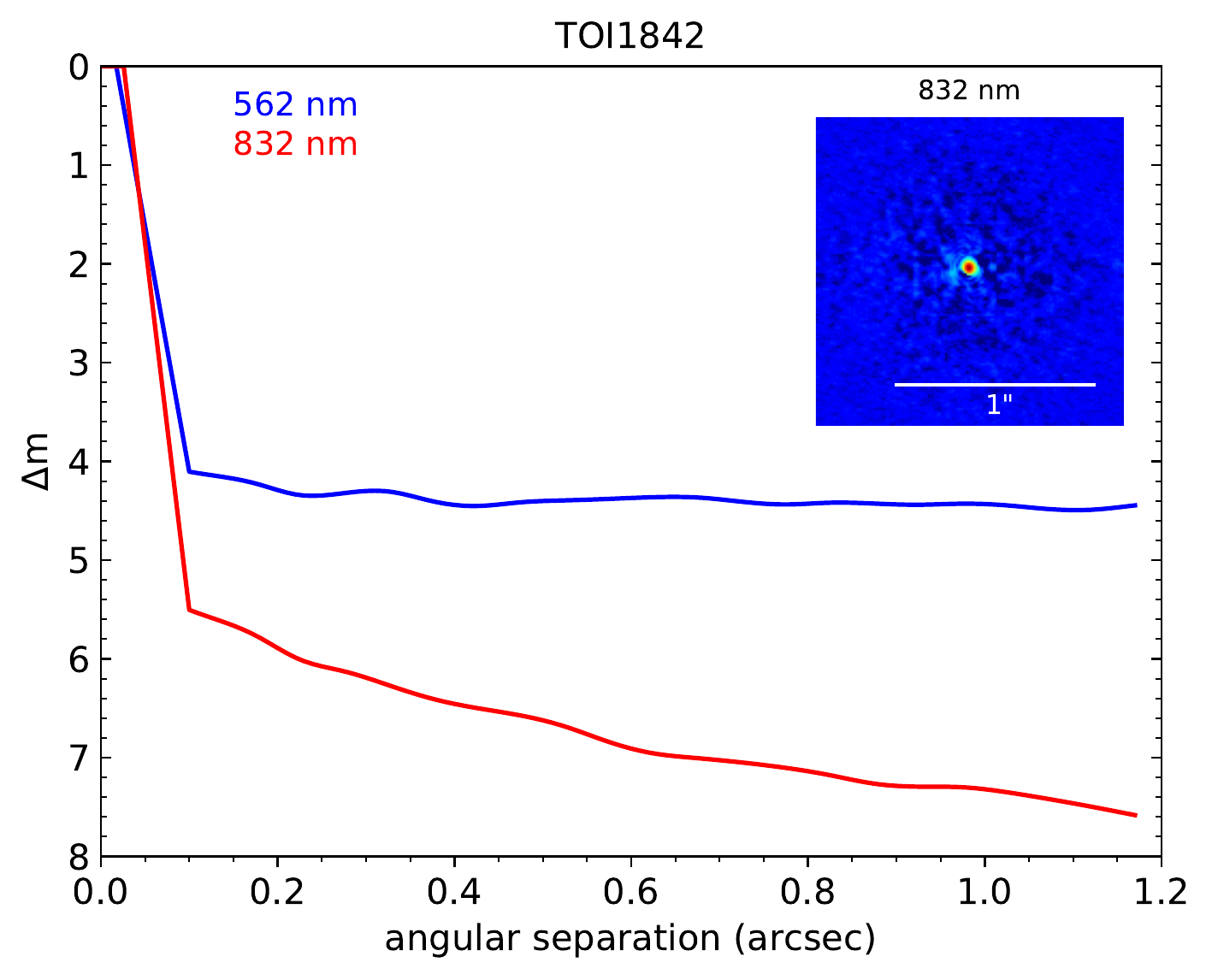}
  \caption{Zorro speckle observations of TOI-1842 obtained at 562 and 832\,nm.  The lines (red: 832\,nm; blue: 562\,nm) represent the $5\sigma$ contrast levels reached, revealing that no companion star is detected from the diffraction limit (17\,mas) out to 1.2\,\arcsec within a $\Delta$\,mag of 4 to 7.5.  The inset shows the 832\,nm reconstructed speckle image and has north up and East to the left.}
  \label{fig:zspeckle}
\end{figure}

We observed TOI-1842 with optical speckle interferometric imaging as part of our standard process for validating transiting exoplanets to assess the possible contamination of bound or unbound companions on the derived planetary radii \citep{ciardi2015}.  Speckle imaging observations of TOI-1842 were performed on 1 March 2021 UT using the Zorro speckle instrument on Gemini-South\footnote{https://www.gemini.edu/sciops/instruments/alopeke-zorro/}.  Zorro simultaneously provides speckle imaging in two bands, 562\,nm and 832\,nm, with output data products including a reconstructed image, and robust limits on companion detections \citep{howell2011}.  Figure~\ref{fig:zspeckle} shows our results for TOI-1842 from both bandpasses and the reconstructed 832\, nm speckle image.  We find that TOI-1842 is indeed a single star with no detected companion brighter than $\Delta_{\textrm{mag}}\,\sim$ 4-7.5 magnitudes (M0V) detected within 1.2\,\arcsec. The angular resolution limits of the speckle observations, at the distance of TOI-1842 (d=223.5 pc), correspond to spatial limits of 4.5 to 268 au.

\section{Analysis and Results}\label{sec:Results}

We used the {\tt Exo-Striker} exoplanet toolbox\footnote{\url{https://github.com/3fon3fonov/exostriker}}
\citep{Trifonov2019_es} to perform a signal analysis and a joint fit of the transit light curve and radial velocity data.  
For the photometric transit period search, {\tt Exo-Striker} 
uses the {\tt transitleastsquares} package \cite[TLS;][]{Hippke2019b}, and for the RV period search analysis, it uses the generalized Lomb-Scargle periodogram \citep[GLS;][]{Zechmeister2009}. 
{\tt Exo-Striker} allows for wide variety of joint-modeling schemes of multi-telescope transit and RV data consistent with planetary signals.
The modeling can be performed either by `best-fit' optimization schemes (i.e. Levenberg-Marquardt, Nelder-Mead, Newton, etc.), 
or sampling schemes such as an affine-invariant ensemble Markov Chain Monte Carlo (MCMC) sampler \citep{Goodman2010} 
via the \texttt{emcee} package \citep{emcee}, and the nested sampling technique \citep{Skilling2004} via {\tt dynesty} sampler \citep{Speagle2020}, which allow detailed posterior probability analysis.

\subsection{Signal search}
\label{sec:tra_search}

\begin{figure}
    \centering
 
    \includegraphics[width=9cm]{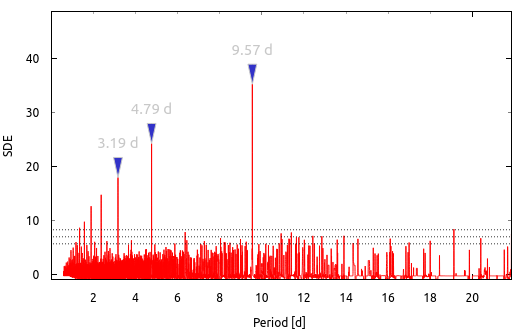}    
     \caption{TLS power spectra of the detrended \textit{TESS} PDCSAP light curve of Sector 23 and LCOGT data.
    Horizontal dashed lines indicates the signal detection efficiency (SDE) power level of 5.7, 7.0, and 8.3, which corresponds to the TLS false positive rate of 10\%, 1\% and 0.1\%.
    The strongest TLS power appears at 9.57 days, the remaining significant peaks are the low-frequency harmonics of the transit signal.
     } 
    \label{TLS} 
\end{figure}

\autoref{TLS} shows the TLS signal detection efficiency (SDE) power spectra of the detrended relative flux of the \textit{TESS} and the LCOGT data of TOI-1842. We detect a significant transit event on the combined light curve data with a period of 9.5742$\pm$0.0024 days, t$_0$ = 2458933.32187,
a mean transit duration of 4.272 h, and a mean transit depth of 0.9967 in relative flux units. 
\autoref{TLS} also shows that TLS detects other significant periodicities in the data, which, however, are attributed to the 1/2, 1/3, 1/4, etc., low-frequency harmonics of the actual transit period.  
Our independent transit search analyses are in agreement with the available SPOC DV estimates based on the \textit{TESS} Sector 23 data of TOI-1842.

\begin{figure}
    \centering

     \includegraphics[width=9cm]{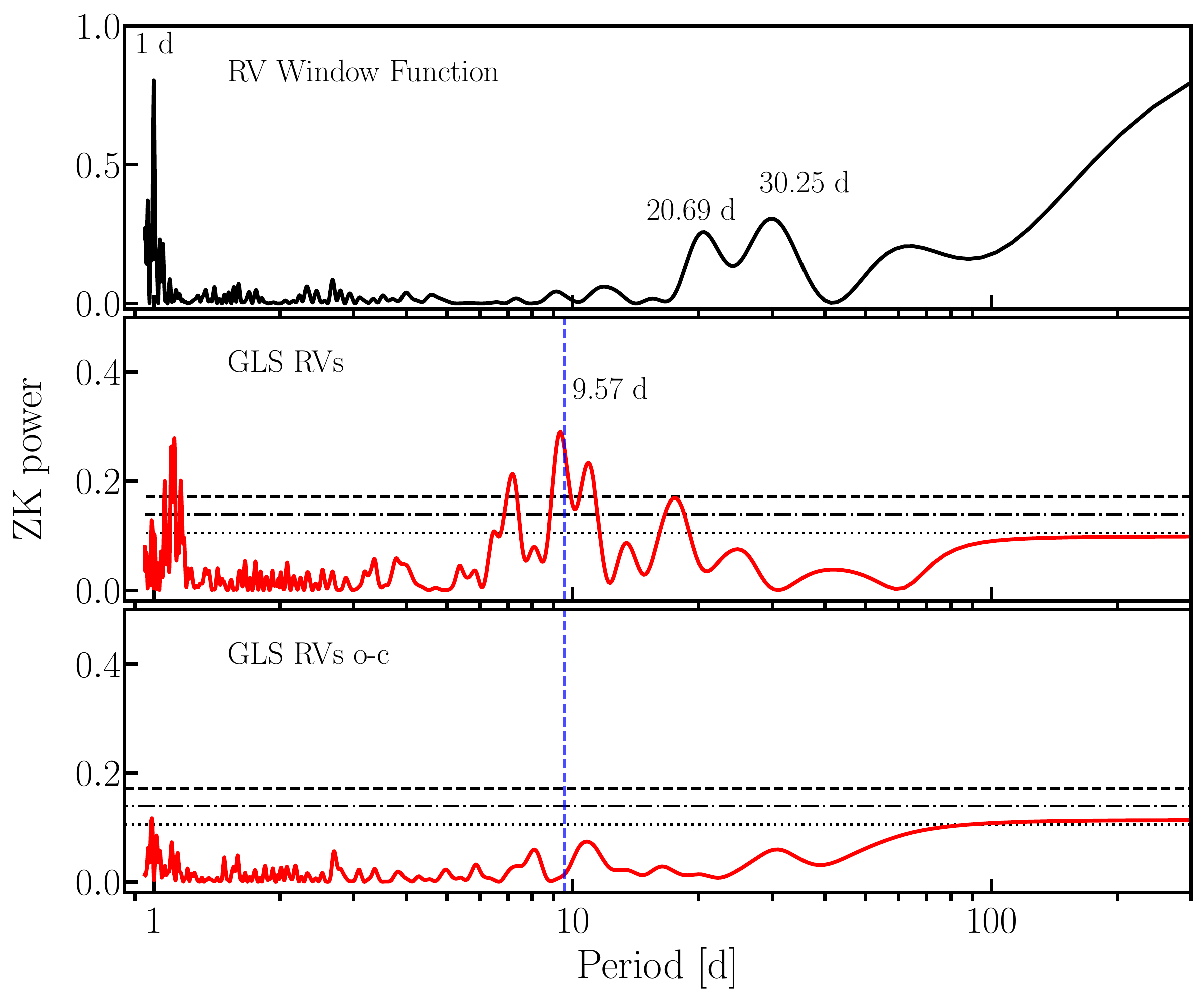}

     \caption{{\em Top panel}: Window function power spectrum of the combined Doppler data for TOI-1842. 
     {\em Middle panel}: GLS power spectra of the Doppler data of TOI-1842. Horizontal dashed lines indicate GLS false positive thresholds of 10\%, 1\% and 0.1\%. A significant periodic signal is detected at a period of  9.34\,d, which is induced by TOI-1842b. The other significant signals near $\sim$10.8\,d, 7.10\,d and 1.12\,d are aliases of the window function and the planetary signal.
     {\em Bottom panel}: Same as above panel, but for the RV residuals of the best joint  model of TOI-1842.
     } 
    \label{GLS} 
\end{figure}

\autoref{GLS} shows the period search analysis of the combined \textsc{Minerva}-Australis and NRES RV data set.
The top panel of  \autoref{GLS} shows the window function of the available Doppler data of TOI-1842. The observational scheduling results in strong window function power at periods of 1 d, 20.54 d, 29.98 d, and lower frequencies.
The middle panel of \autoref{GLS} shows the GLS power spectra of the Doppler data of TOI-1842. Horizontal dashed lines indicate GLS false-positive thresholds of 10\%, 1\%, and 0.1\%. We detect a significant periodic signal with a period of  9.34 $\pm$ 0.15 d, which is near the period of the transit signal. The remaining significant signals are aliases of the window function and the Doppler-induced planetary signal. The bottom panel of \autoref{GLS} shows the GLS power spectrum of the RV residuals of the joint fit (see \autoref{sec:joint}).  We did not detect any significant residual RV periodicity suggesting the presence of any additional planet.

\begin{figure*}
    \centering

    \includegraphics[width=8.8cm]{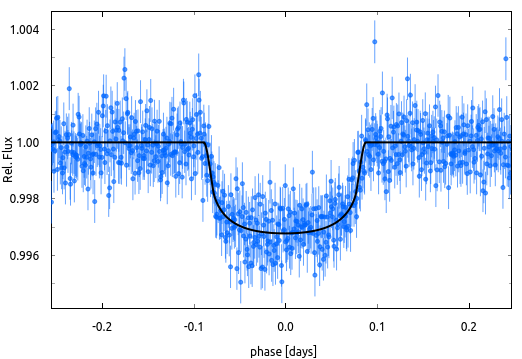}
    \includegraphics[width=8.8cm]{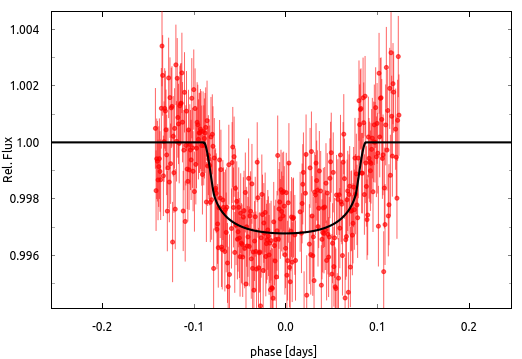}\\
    \includegraphics[width=8.8cm]{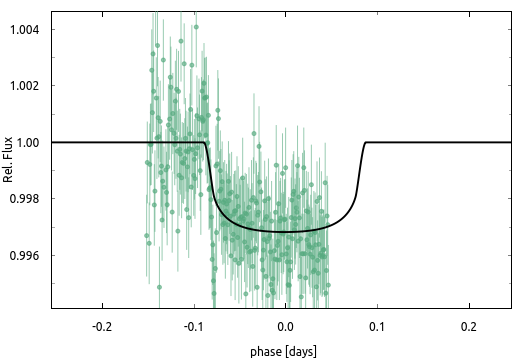}
    \includegraphics[width=8.8cm]{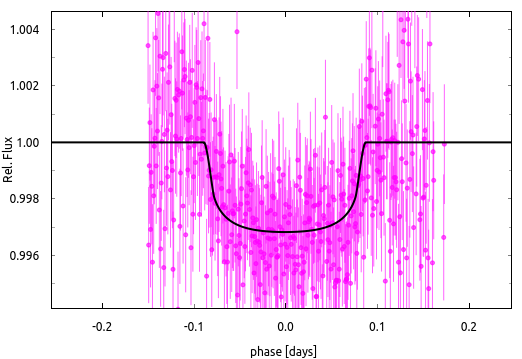}\\

     \caption{Transit photometry data and model fit for {\it TESS} (blue, Sector 23), LCOGT data from SAAO (red), and CTIO (green and magenta).  
     The transit model (black solid line) is constructed jointly with precise Doppler data as described in \autoref{sec:joint}.
     } 
    \label{Transit_plots} 
\end{figure*}

\begin{figure*}
    \centering
 
    \includegraphics[width=18cm]{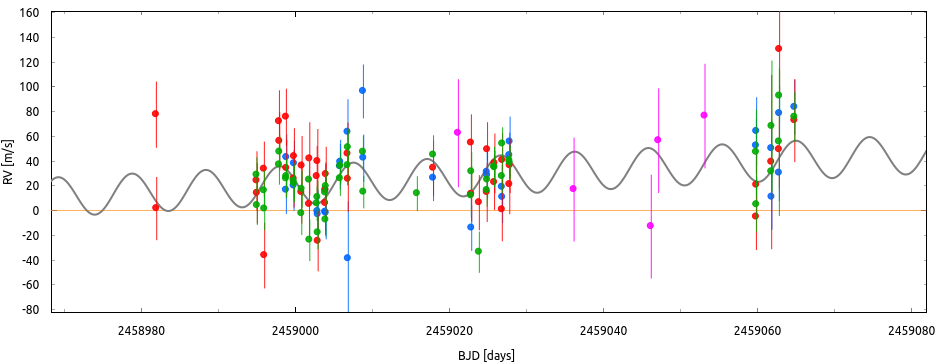}\\
     \includegraphics[width=8.8cm]{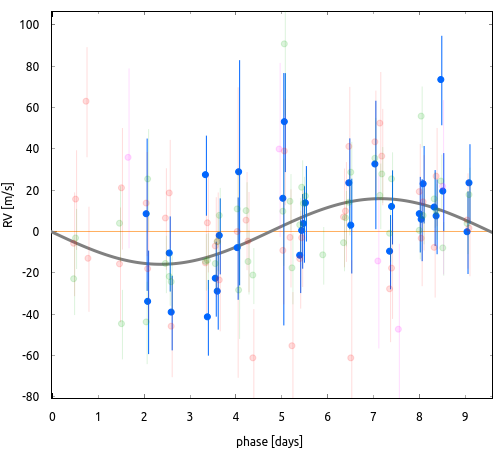}
     \includegraphics[width=8.8cm]{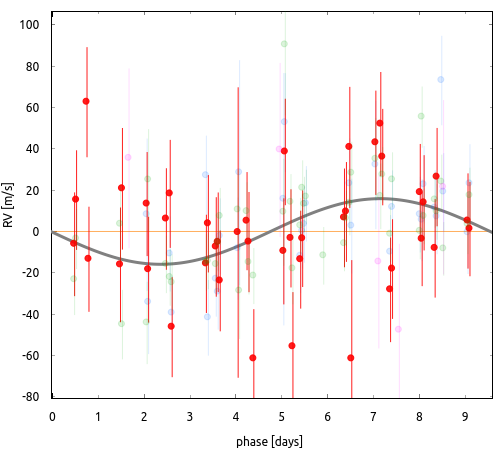}\\
     \includegraphics[width=8.8cm]{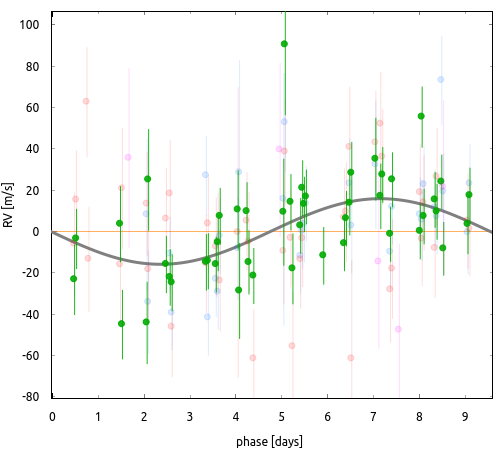}
     \includegraphics[width=8.8cm]{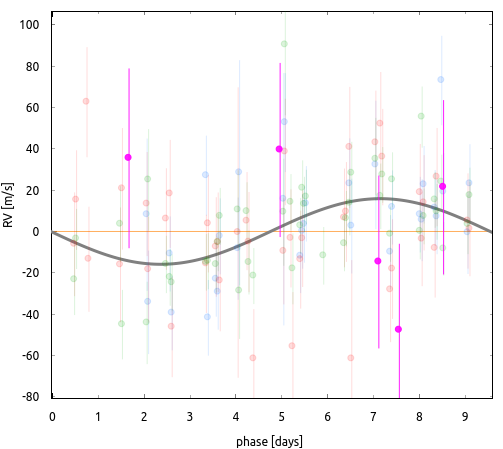}\\

     \caption{Radial velocity data and model fit for TOI-1842.  \textit{Top:} complete time series and joint fit model including a linear trend.  \textit{Bottom:} Phase-folded RV data, with each of the four instruments individually highlighted (with heavier points) in each subpanel for clarity.  Left to right subpanels: \textsc{Minerva}-Australis Telescopes 1 (blue), 4 (red), 5 (green), and NRES/SAAO (purple).
     } 
    \label{RV_plots} 
\end{figure*}

\subsection{Combined transit and RV orbital analysis}
\label{sec:joint}

We use the TLS and GLS results from \autoref{sec:tra_search} as a starting point for our combined transit and RV orbital modeling.
From these estimates and the following intermediate transit and RV parameter optimization analysis with the {\tt Exo-Striker}, we define informative and non-informative parameter priors needed for our global parameter search.
We performed multi-dimensional parameter fitting, constructing the posterior probability distribution by employing a nested sampling scheme using the {\tt dynesty} sampler. 
We set  100 ``live-points'' per fitted parameter, sampled via random walk, dynamic nested sampling with a 100\% weight on the posterior convergence \citep[see,][for details]{Speagle2020}. 
Our adopted dynamic nested sampling scheme allows us to construct a posterior probability distribution, from which we adopted the 68.3\,\% confidence levels as $1\sigma$ parameter uncertainties.
In addition, the nested sampling analysis is very convenient for an adequate model comparison via the difference of their Bayesian log-evidence $\Delta\ln Z. $\footnote{$\Delta\ln Z = \ln Z_{\rm complex~model} -\ln Z_{\rm simpler~model}$. Two models are indistinguishable if their Bayesian log-evidence difference satisfies $\Delta \ln Z \lesssim 2$, a model moderately favored over another if $\Delta\ln Z$ $>$ 2, and strongly favored if $\Delta \ln Z > 5$ \citep{Trotta2008}.}
Our adopted parameter priors, the resulting median values and their 1$\sigma$ uncertainties drawn from the posterior probability distribution, and the sample with the maximum $-\ln{\mathcal L}$  (i.e., best-fit) for our final orbital analysis are summarized in \autoref{NS_param}. \autoref{Transit_plots} 
shows the best-fit transit model component applied to Sector 21 and the LCOGT data, whereas \autoref{RV_plots} shows the RV component of the joint-fit model applied to the combined RVs. 
Our combined transit and RV orbital analyses are performed as follows:

The transit modeling within the {\tt Exo-Striker} is done by a wrapper of the BAsic Transit Model cAlculatioN package \citep[{\tt batman}, ][]{Kreidberg2015}. 
This package constructs the transit light curve model by accepting planetary orbital parameters such as; period $P$, eccentricity $e$, the argument of periastron, $\omega$, and inclination $i$, which are common to the Keplerian RV model of the {\tt Exo-Striker}. Additional transit parameters of {\tt batman} are the time of inferior transit conjunction $t_{\rm 0}$, and the planet semi-major axis and radius in units relative to the stellar radius $a$/$R_\star$ and $R$/$R_\star$, respectively. The RV model component fits the RV signal semi-amplitude $K$, which constrains the planetary mass.

In addition to the orbital parameters, our modeling scheme includes a number of nuisance parameters, which we optimize simultaneously.
For each transit and RV data set we  
 fit their respective relative mean offset and additional data ``jitter'' terms, the latter of which was simultaneously added in quadrature to the data error budget while evaluating the models' (negative) log-likelihood $-\ln\mathcal{L}$ \citep{Baluev2009}. In particular, we model the light curve of $\tess$ obtained in Sector 21 and the LCOGT data of SAAO, CTIO 1, and CTIO 2, which adds eight additional parameters. We adopt separate quadratic limb-darkening transit models  of $\tess$ and LCOGT, for which we optimize separate quadratic limb-darkening coefficients, $u1_{\rm TESS}$ and $u2_{\rm TESS}$, and $u1_{\rm LCOGT}$ and $u2_{\rm LCOGT}$, respectively. 
  For the RV data, we fit the precise RVs of \textsc{Minerva} obtained with telescopes 1, 4, and 5 and the NRES-CPT, which also adds eight nuisance parameters. We do not include the NRES-Wise RVs in our joint fit since we found that their scatter is too large, and overall, does not improve the outcome of our analysis.
 In addition, we find a marginally significant positive Doppler drift in the combined RV data, which is possibly induced by another companion orbiting at a greater distance.  No companions are visible (Figure 2), and there is only one object within 1 arcmin, a faint background star at $T=17.595$, separated by 28 arcsec.  Since the nature of this signal cannot be determined based on the available data, we model it by applying an additional RV linear trend parameter.


We performed two sets of joint fit nested sampling analysis; for a forced  
circular orbit of TOI-1842\,b ($e$=0, for which $\omega$ is undefined),
and with free $e$ and $\omega$.
We found that the simpler circular model is moderately favored with a log-evidence of $\ln Z = 82096.06$, whereas the full-Keplerian model has a slightly poorer log-evidence of $\ln Z = 82093.65$.  From the full-Keplerian model, we find that the orbital eccentricity is consistent with 0, but with large uncertainties, from which we estimate an upper eccentricity limit of $e$ $<$ 0.25 at the 1$\sigma$ level.
Thus, as a final orbital solution of TOI-1842\,b, we adopted 
the simpler, circular orbit joint fit model, which is presented in  \autoref{NS_param}, \autoref{Transit_plots}, and \autoref{RV_plots}.

The values of three most representative parameters from our best-fit are 
$P =   9.5739_{-0.0001}^{+0.0002}$\,d,
 $t_{0} = 2459402.4575_{-0.0024}^{+0.0026}$\,d, 
 and $K = 15.9_{-2.80}^{+2.9}$ m\,s$^{-1}$.
Given the parameter posteriors, and our estimates of the stellar mass, radius and their uncertainties we derive a dynamical planetary mass of 
$m_{\rm pl.}$ = 0.214$_{-0.038}^{+0.040}$ $M_{\rm Jup}$,
semi-major axis of $a = 0.1001_{-0.0007}^{+0.0007}$ au, and a planetary radius of $R_{\rm pl.} = 1.04_{-0.05}^{+0.06} R_{\rm Jup}$, which makes it one of the lowest density planets discovered to date ($\rho$=0.252$\pm$0.091 g cm$^{-3}$).

We further inspected the {\it TESS} residuals with TLS and the RV residuals with GLS to check for evidence of additional planets, but none was found.

\section{Discussion \& Conclusions} \label{sec:Discussion}

We have confirmed the planetary nature of the warm Saturn TOI-1842b ($P=9.5737\pm$0.0015 d), and we have measured its mass to be 0.214$^{+0.040}_{-0.038}$ \mj, with a radius of 1.04$^{+0.06}_{-0.05}$ \rj.  This planet, moving on a circular orbit with $a=0.10$\,au\ about its subgiant host star, joins a small cohort of 56 confirmed planets orbiting within 0.5\,au of evolved stars.  With a period of nearly 10 days, this planet's orbital properties are similar to TOI-481b \citep{toi481}, another warm giant ($T_{eq}=1210\pm$29\,K) orbiting a slightly evolved metal-rich star.  While short-period giant planets orbiting stars more massive than the Sun remain rare, such planets appear to be preferentially found around super-Solar metallicity stars, as evidenced by the positive planet-metallicity correlation found for evolved stars by several studies \citep[e.g.][]{reffert15, jones16, ppps8, wolthoff21}.  \citet{petigura18} also found that warm Saturns in particular are intrinsically less common than close-in Neptunes or Jupiters, and tend to be hosted by metal-rich stars.  The properties of TOI-1842b are thus consistent with this class of rare low-mass gas giants. 

\begin{figure}[!ht]
    \centering
    \includegraphics[width=\columnwidth]{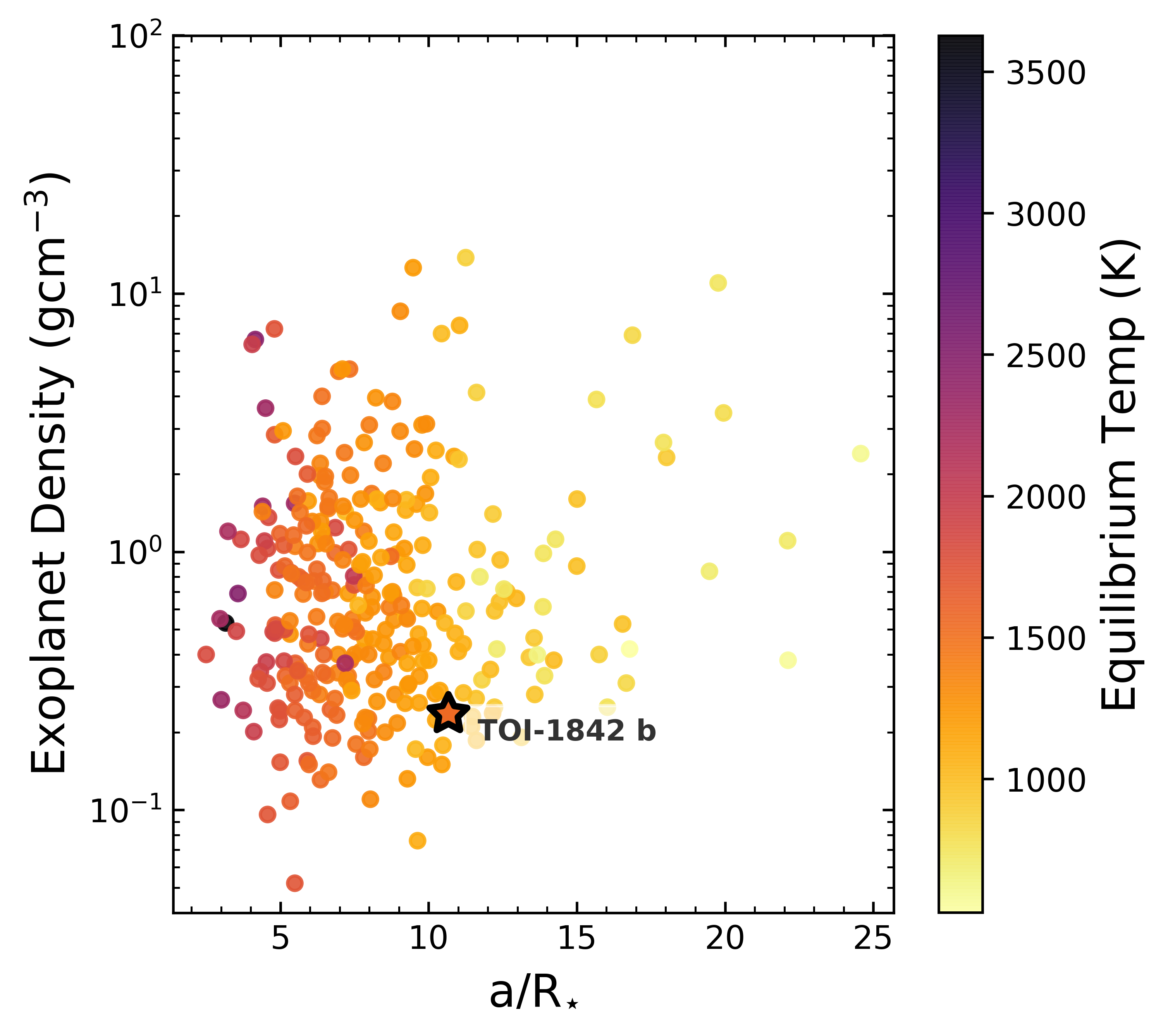}
\caption{Planetary density as a function of orbital distance $a/R_{*}$ for 410 Hot-Jupiters.  Only 37 Hot Jupiters have bulk densities smaller than TOI-1842b (starred in this plot).  \label{fig:density}}
\end{figure}
\begin{figure}[!ht]
    \centering
    \includegraphics[width=\columnwidth]{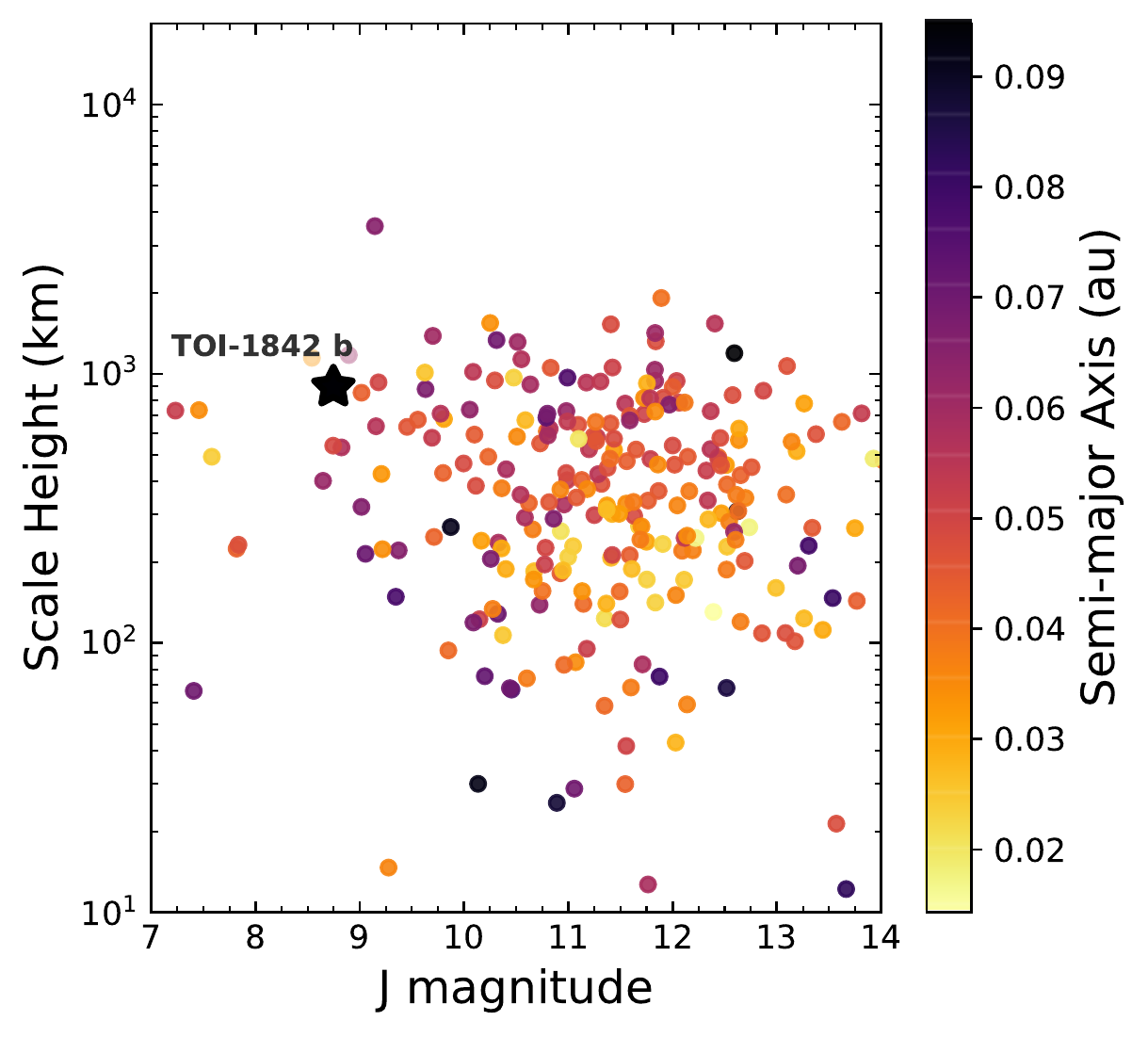}
\caption{Estimated atmospheric scale height versus $J$ magnitude for 410 giant planets ($R_p>0.75$\rj).  With a scale height of 893 km, TOI-1842b is an attractive target for atmospheric characterisation.  Only WASP-76b \citep{west16} presents a larger scale height and a brighter host star. \label{fig:thicc}}
\end{figure}

TOI-1842b is remarkable due to its low density ($\rho$=0.252$\pm$0.091 g cm$^{-3}$), which ranks it among the least-dense 9.3\% of known planets.  Figure~\ref{fig:density} places this planet in context of the cohort of planets with measured bulk densities.  From the derived equilibrium temperature (1210\,K) and surface gravity of the planet and assuming a H/He atmosphere with a mean molecular mass of $\mu=2.3$\,amu, the calculated atmospheric scale height (from $H_{\mathrm{b}}=kT_{eq}/(\mu g_{\mathrm{b}})$) is $H_{\mathrm{b}}=893$\,km.\footnote{For comparison, Jupiter has the largest scale height in our Solar System at 27 km.}  The large scale height of TOI-1842b ($h/R_p$=1.2\%) is reminiscent of KELT-11b (Figure~\ref{fig:thicc}).  While KELT-11b has a larger scale height \citep{kelt11}, its primary transit time of $\sim8$ hours would require multiple full night observations to fully observe.  Alternatively, TOI-1842b has a transit duration of $\sim4$ hours, making it an ideal candidate for single night observation follow up.  The transmission spectroscopy metric \citep[TSM, see,][]{kempton18}, used to assess the suitability of transmission spectroscopy observations with the James Webb Space Telescope (\textit{JWST}), for TOI-1842\,b is 141.  Planets with TSM values greater than $\sim$96 (for Jovians and sub-Jovians), such as for TOI-1842b, are considered suitable for these observations with \textit{JWST}.

  The measured mass and radius of TOI-1842b place it within the category of ``inflated sub-Saturns.''  From the relationship given in \citet{weiss13}, TOI-1842b has a predicted radius of 9\re; its measured radius of 11.65\re\ is thus $\sim$30\% larger than a non-inflated planet with its mass and incident flux.  Inflated sub-Saturn planets occupy a relatively unexplored parameter space that is seen as a key transitional population group between inflated super-Earths and Jovian worlds~\citep{LeeChiang2016, colon2020}. Conversely, the low density of some planets in longer period orbits, like TOI-1842b, might be indicative of their formation pathways. ~\citet{LeeChiang2016} proposed that low density sub-Saturns formed in-situ.  The chemical composition of these planets should reflect the composition of the gas disk from which they were formed~\citep{Oberg2011}.  Furthermore, TOI-1842b is a rare close-in planet orbiting an evolved star.  By observing and identifying the molecular species and physical dynamics present within the atmosphere of TOI-1842b, we can obtain a better understanding on how these inflated gas giants interact with their host star over time and better determine the formation process for this population.

The radius and mass of TOI-1842b, at its orbital distance of 0.1\,au, are consistent with a coreless planet (cf. Table 4 in \citealt{fortney07}).  We note that the host star is metal-rich with nearly twice Solar [M/H], making the presence of a coreless giant planet most intriguing.  One would expect metal-rich protoplanetary disks to form planets with larger solid cores due to a higher surface density \citep{pollack1996}.  For example, HD\,149026 with a similar metal content hosts a planet with a $\sim$67\me\ core \citep{sato2005}.  TOI-1842b thus presents a fascinating conundrum; a relatively rare class of planet for which detailed atmospheric studies will be eminently feasible and illuminating.  



\begin{table*}[ht]

\centering   
\caption{{Nested sampling posteriors and maximum $-\ln\mathcal{L}$ of the orbital and 
nuisance parameters of the TOI-1842 system, derived by 
joint modeling of photometry ($\tess$, LCO) and radial velocities (MINERVA, NRES).
}}
\label{NS_param}


\begin{tabular}{p{6.95cm} r r r r r p{6.95cm} p{7.95cm} p{3.95cm} p{3.95cm} p{3.95cm} p{3.95cm} rrrrrrr }     

\hline\hline  \noalign{\vskip 0.7mm}


  Parameter  &  Median and $1\sigma$  & Max. $-\ln\mathcal{L}$  & Adopted priors  \\ \noalign{\vskip 0.9mm}
\hline

$K$  [m\,s$^{-1}$]            & 15.9$_{-2.80}^{+2.9}$ 
                              & 16.8  
                              &  $\mathcal{U}$(5.0,30.00)   \\ \noalign{\vskip 0.9mm}

$P$  [day]                    &  9.5739$_{-0.0001}^{+0.0002}$  
                              & 9.5739  
                              & $\mathcal{U}$(9.565,9.580)    \\ \noalign{\vskip 0.9mm}



$t_{\rm 0}$  [deg]            & 2459402.4575$_{-0.0024}^{+0.0026}$  
                              & 2459402.4572   
                              & $\mathcal{U}$(2459402.2,2459402.6) \\ \noalign{\vskip 0.9mm}
                              
$i$          [deg]            & 88.4$_{-1.5}^{+0.9}$  
                              & 88.3  
                              & $\mathcal{U}$(80.0,90.0) \\ \noalign{\vskip 0.9mm}  
                              

a/$R_\star$                   &  16.2$_{-2.9}^{+1.6}$   
                              & 16.1
                              &  $\mathcal{U}$(5.0,15.00)   \\ \noalign{\vskip 0.9mm}
                              
R/$R_\star$                   &  0.052$_{-0.002}^{+0.003}$  
                              &  0.052  
                              &  $\mathcal{U}$(0.01,0.10)  \\ \noalign{\vskip 1.9mm}


 $a$  [au]                     &  0.1001$_{-0.0007}^{+0.0007}$ 
                              &  0.0925    
                              &  (derived)   \\ \noalign{\vskip 0.9mm}

$m_{p}$  [$M_{\rm jup}$]          &  0.214$_{-0.038}^{+0.040}$  
                              & 0.225   
                              &  (derived)   \\ \noalign{\vskip 0.9mm}

$R_p$  [$R_{\rm jup}$]          &  1.04$_{-0.05}^{+0.06}$ 
                              &  1.05 
                              &  (derived)  \\ \noalign{\vskip 1.9mm}                           


RV off.$_{\rm MINERVA~TEL~1}$ [m\,s$^{-1}$] &  3603.0$_{-11.6}^{+11.2}$ 
                                      &  3595.3 
                                      &  $\mathcal{U}$(3200,3800)    \\ \noalign{\vskip 0.9mm} 
                                      
RV off.$_{\rm MINERVA~TEL~4}$ [m\,s$^{-1}$] &  3652.5$_{-11.4}^{+10.8}$ 
                                      &  3646.0  
                                      &  $\mathcal{U}$(3200,3800)    \\ \noalign{\vskip 0.9mm} 
                                      
RV off.$_{\rm MINERVA~TEL~5}$ [m\,s$^{-1}$] &  3665.9$_{-10.5}^{+9.9}$ 
                                      &  3659.3 
                                      &  $\mathcal{U}$(3200,3800)    \\ \noalign{\vskip 0.9mm}       
                                      
RV off.$_{\rm NRES}$ [m\,s$^{-1}$]  &  3337.5$_{-28.3}^{+34.9}$  
                                      &  3332.9 
                                      &  $\mathcal{U}$(3200,3800)    \\ \noalign{\vskip 0.9mm}          
                                      

RV jitter$_{\rm MINERVA~TEL~1}$ [m\,s$^{-1}$] &  15.2$_{-4.6}^{+3.9}$    
                                      &  15.7     
                                      &  $\mathcal{J}$(1.00,25.00)    \\ \noalign{\vskip 0.9mm} 
                                      
RV jitter$_{\rm MINERVA~TEL~4}$ [m\,s$^{-1}$] &  18.7$_{-5.8}^{+3.4}$    
                                      &  17.5     
                                      &  $\mathcal{J}$(1.00,25.00)    \\ \noalign{\vskip 0.9mm} 
                                      
RV jitter$_{\rm MINERVA~TEL~5}$ [m\,s$^{-1}$] &  9.5$_{-3.5}^{+3.3}$    
                                      &  8.8  
                                      &  $\mathcal{J}$(1.00,25.00)    \\ \noalign{\vskip 0.9mm} 
                                      
RV jitter$_{\rm NRES}$ [m\,s$^{-1}$]  &  43.2$_{-16.6}^{+22.1}$     
                                      &  45.4      
                                      &  $\mathcal{J}$(1.00,100.00)    \\ \noalign{\vskip 0.9mm}     
  

RV lin.\,trend [m\,s$^{-1}$/day]                   &  0.316$_{-0.121}^{+0.254}$     
                                      &  0.395      
                                      &  $\mathcal{J}$(-5.00,5.00)    \\ \noalign{\vskip 1.9mm}     


Transit offset$_{\rm TESS-S21}$ [ppm] &  46$_{-8}^{+8}$  
                                      &  45  
                                      &  $\mathcal{N}$(0.0,1000.0)     \\ \noalign{\vskip 0.9mm}
 
Transit offset$_{\rm LCO-SAAO}$ [ppm]     &  11$_{-108}^{+110}$  
                                     &  $-$5 
                                     &  $\mathcal{N}$(0.0,1000.0)     \\ \noalign{\vskip 0.9mm} 
  
Transit offset$_{\rm LCO-CTIO~1}$ [ppm]     &  158$_{-127}^{+140}$  
                                     &  118 
                                     &  $\mathcal{N}$(0.0,1000.0)     \\ \noalign{\vskip 0.9mm} 
  
Transit offset$_{\rm LCO-CTIO~2}$ [ppm]     &  $-$1834$_{-122}^{+132}$  
                                     &  $-$1811
                                     &  $\mathcal{N}$(0.0,1000.0)     \\ \noalign{\vskip 0.9mm}


Transit jitter$_{\rm TESS-S21}$ [ppm] &  40$_{-23}^{+50}$ 
                                      &  22 
                                      &  $\mathcal{J}$(0.0,1000.0) \\ \noalign{\vskip 0.9mm}   
                                     
Transit jitter$_{\rm LCO-SAAO}$ [ppm]   & 1143$_{-82}^{+94}$  
                                    & 1127
                                    &  $\mathcal{J}$(0.0,1000.0) \\ \noalign{\vskip 1.9mm} 
                                    
Transit jitter$_{\rm LCO-CTIO~1}$  [ppm]   & 1275$_{-85}^{+123}$  
                                    & 1277 
                                    &  $\mathcal{J}$(0.0,1000.0) \\ \noalign{\vskip 1.9mm} 
                                    
 Transit jitter$_{\rm LCO-CTIO~2}$  [ppm]   & 1638$_{-85}^{+120}$  
                                    & 1635 
                                    &  $\mathcal{J}$(0.0,1000.0) \\ \noalign{\vskip 1.9mm}


Quad. limb-dark.$_{\rm TESS}$ $u_1$    &   0.307$_{-0.175}^{+0.235}$  
                                   &   0.151   
                                   &  $\mathcal{U}$(0.00,1.00)   \\ \noalign{\vskip 0.9mm}
                                   
Quad. limb-dark.$_{\rm TESS}$ $u_2$    &  0.586$_{-0.315}^{+0.255}$  
                                   &  0.718 
                                   &  $\mathcal{U}$(0.00,1.00)   \\ \noalign{\vskip 0.9mm} 

Quad. limb-dark.$_{\rm LCO}$ $u_1$  &  0.327$_{-0.190}^{+0.226}$  
                                    &  0.292 
                                    &  $\mathcal{U}$(0.00,1.00)    \\ \noalign{\vskip 0.9mm} 
                                    
Quad. limb-dark.$_{\rm LCO}$ $u_2$  &  0.481$_{-0.291}^{+0.293}$  
                                    &  0.508 
                                    &  $\mathcal{U}$(0.00,1.00)   \\ \noalign{\vskip 0.9mm}


\hline \noalign{\vskip 0.7mm}

\end{tabular}


\end{table*}

\begin{figure}[!ht]
    \centering
\includegraphics[width=0.8\columnwidth]{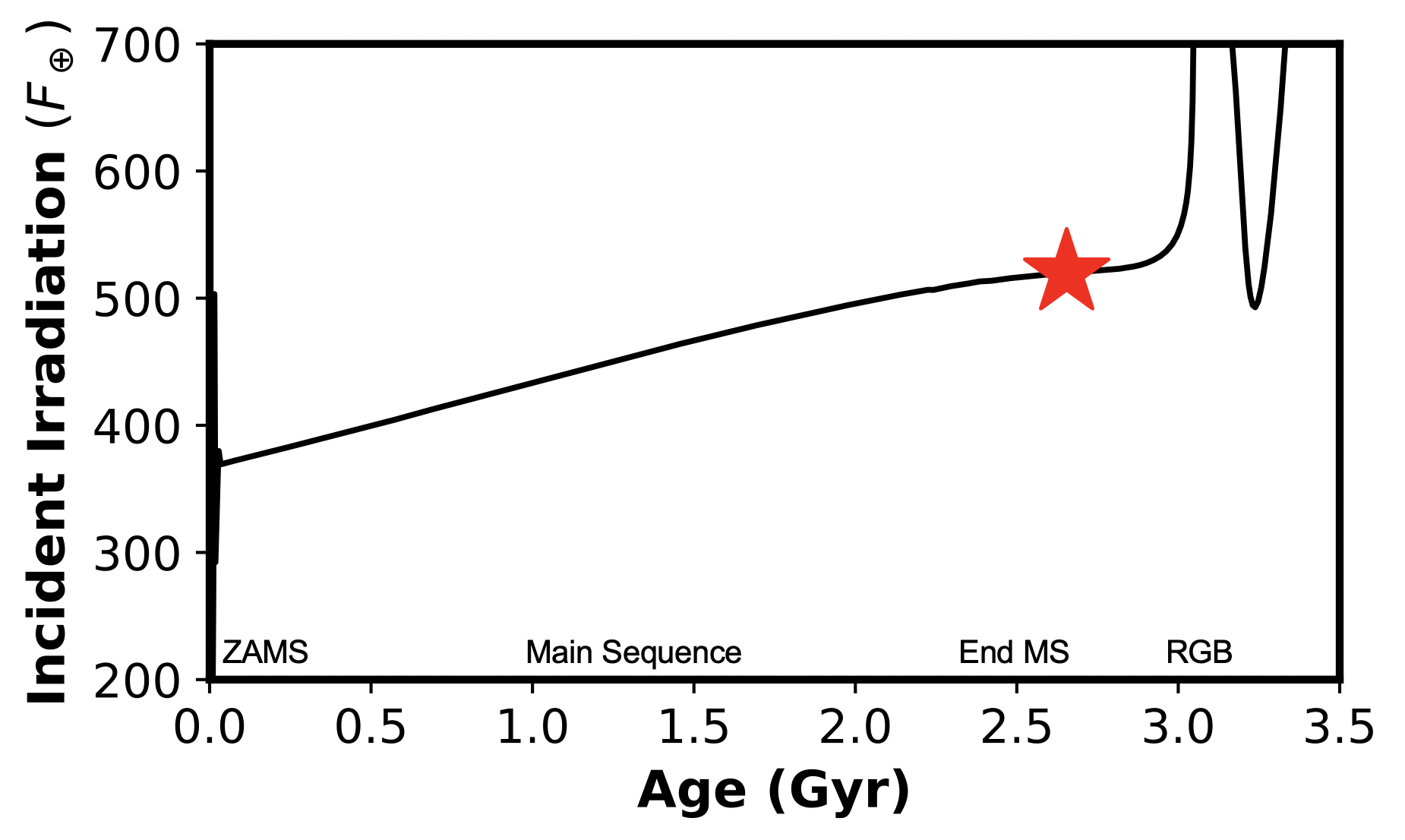}
\caption{Incident irradiation received by TOI-1842b over the course of the  host star evolution. Assuming no tidal modification of the orbit of the planet, TOI-1842b is now receiving 30\% more incident irradiation ($\sim$550\,F$_{\oplus}$) than it did when the host star settled on the main sequence.\label{fig:fincident}}
\end{figure}

TOI-1842b joins a growing list of low density close-in gas giants orbiting stars evolving off the main sequence \citep[e.g.][]{songhu, ngts12, Sha21}, and provides a suitable test to models that explain the higher-than-expected radii of these gas giants via re-inflation mechanisms \citep[e.g.][]{LopezFortney2016,Komacek2020,Thorngren2021}.  Adopting the stellar parameters from Table~\ref{tab:star} and evolution tracks from the MIST models \citep{Dotter2016}, the current irradiation received by TOI-1842b is 30\% more than it originally received at zero-age main sequence (Figure~\ref{fig:fincident}).  A number of similarly inflated warm Jovian planets have recently been found around post main-sequence stars receiving many times their main-sequence incident irradiation (e.g. \citealt{Hartman2016,Grunblatt2016,kelt11,Grunblatt2017}).  

\begin{figure}[!ht]
    \centering
    \includegraphics[width=0.8\columnwidth]{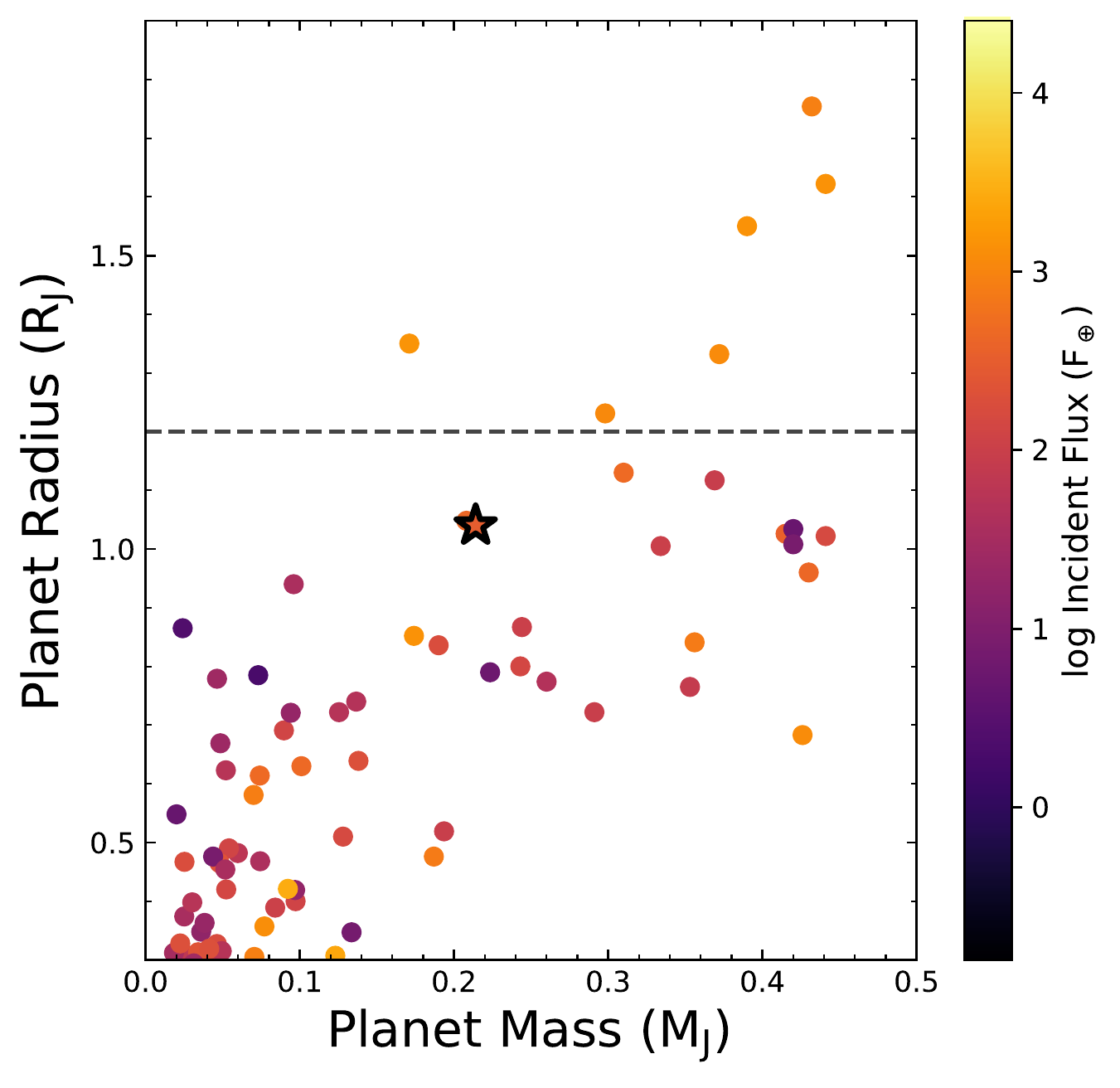}
\caption{Planet radius versus mass (in Jupiter units) for well-characterised Saturns, color-coded by incident flux in terms of the flux received by Earth ($F_{\oplus}$).  TOI-1842b is marked as a large star.  The dashed line at 1.2\rj\ denotes the approximate maximum radius for uninflated planets, as given by \citet{LopezFortney2016}.   \label{fig:fincident2}}
\end{figure}

\begin{figure}[!ht]
    \centering
    \includegraphics[width=0.8\columnwidth]{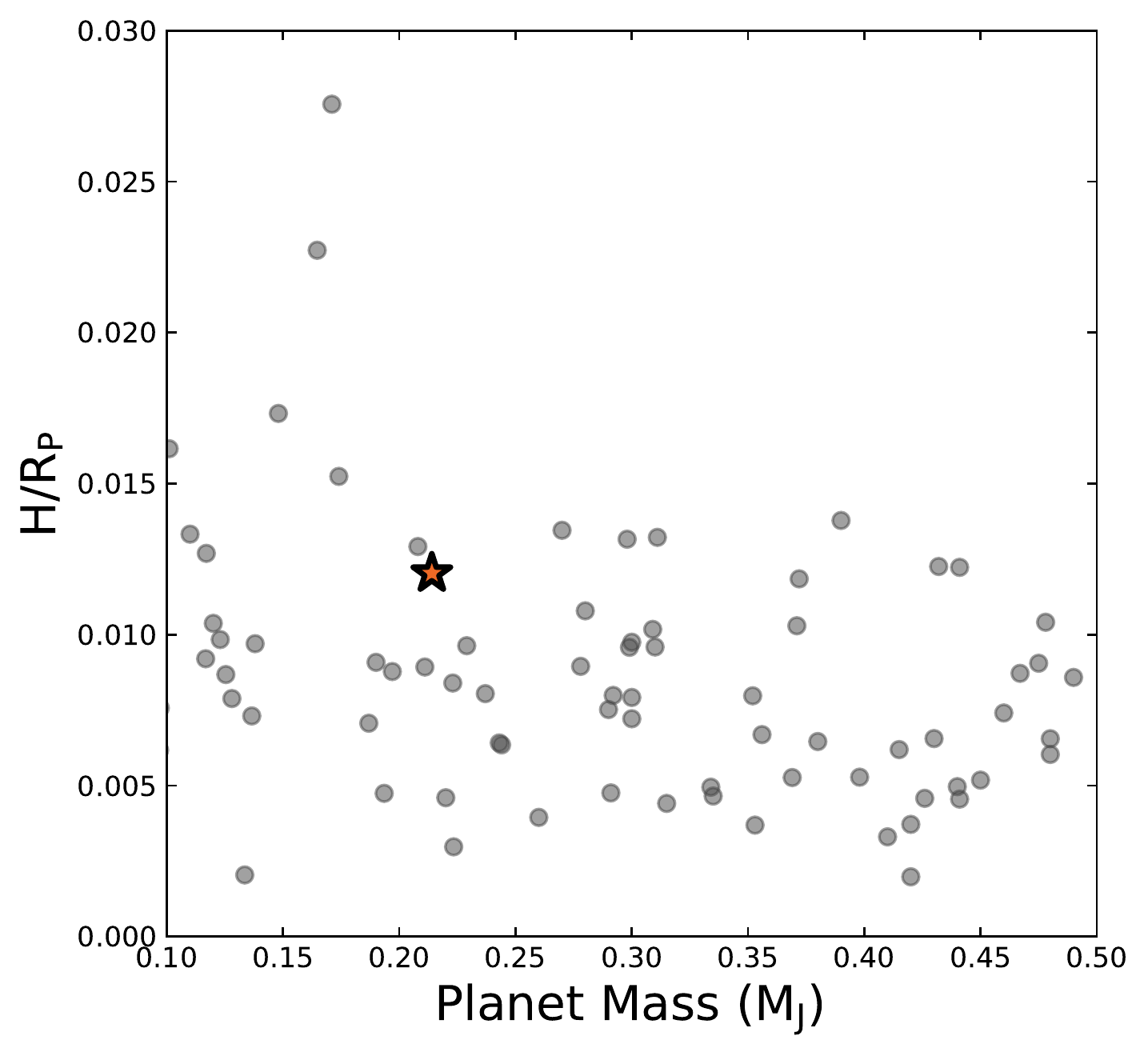}
\caption{Relative atmospheric scale height ($H/R_p$) versus planet mass for well-characterised Saturns.  TOI-1842b is marked as a large star, and is among the most favourable targets in this mass regime.   \label{fig:thicc2}}
\end{figure}

Figure~\ref{fig:fincident2} places TOI-1842b in context with other well-characterised sub-Jovian planets with masses and radii measured to precisions of better than 20\% and 10\%, respectively.  The radius of TOI-1842b is near the upper envelope of similar planets in this mass regime.  TOI-1842 is beginning to leave the main sequence and, as shown in Figure~\ref{fig:fincident}, will irradiate and further inflate TOI-1842b over the next tens of Myr before eventually consuming the planet \citep{villaver14}.  As the star evolves off the main sequence, the strengthening incident irradiation and star-planet tidal interactions may induce increased heating of the planet's core, potentially re-inflating the low density gas giants.  While TOI-1842b is not at present immensely inflated by comparison to planets of similar mass, it stands out in its suitability for atmospheric characterisation.  In addition to the brightness of the host star (cf. Figure~\ref{fig:thicc}), TOI-1842b has a high ratio of atmospheric scale height to planetary radius.  Figure~\ref{fig:thicc2} shows the ratio of atmospheric scale height to planet radius ($H/R_p$) as a function of mass for the cohort of well-characterised sub-Jovian planets.  TOI-1842b lies in the best 10\% of these planets in terms of $H/R_p$.  The two highest points in Figure~\ref{fig:thicc2} with $H/R_p > 0.02$ are KELT-11b \citep{kelt11} and WASP-127b \citep{lam17}.  Those two extremely inflated planets both orbit subgiant stars, and represent possible futures for TOI-1842b as its host star inexorably ascends the subgiant branch.


\acknowledgments
We respectfully acknowledge the traditional custodians of all lands throughout Australia, and recognise their continued cultural and spiritual connection to the land, waterways, cosmos, and community. We pay our deepest respects to all Elders, ancestors and descendants of the Giabal, Jarowair, and Kambuwal nations, upon whose lands the {\textsc{Minerva}}-Australis facility at Mt Kent is situated.

{\textsc{Minerva}}-Australis is supported by Australian Research Council LIEF Grant LE160100001, Discovery Grant DP180100972, Mount Cuba Astronomical Foundation, and institutional partners University of Southern Queensland, UNSW Sydney, MIT, Nanjing University, George Mason University, University of Louisville, University of California Riverside, University of Florida, and The University of Texas at Austin.

Some of the observations in the paper made use of the High-Resolution Imaging instrument Zorro obtained under Gemini LLP Proposal Number: GN/S-2021A-LP-105. Zorro was funded by the NASA Exoplanet Exploration Program and built at the NASA Ames Research Center by Steve B. Howell, Nic Scott, Elliott P. Horch, and Emmett Quigley. Zorro was mounted on the Gemini North (and/or South) telescope of the international Gemini Observatory, a program of NSF’s OIR Lab, which is managed by the Association of Universities for Research in Astronomy (AURA) under a cooperative agreement with the National Science Foundation. on behalf of the Gemini partnership: the National Science Foundation (United States), National Research Council (Canada), Agencia Nacional de Investigación y Desarrollo (Chile), Ministerio de Ciencia, Tecnología e Innovación (Argentina), Ministério da Ciência, Tecnologia, Inovações e Comunicações (Brazil), and Korea Astronomy and Space Science Institute (Republic of Korea).

This work makes use of observations from the LCOGT network.

B.A. is supported by Australian Research Council Discovery Grant DP180100972. 

Funding for the \textit{TESS} mission is provided by NASA's Science Mission directorate.  We acknowledge the use of public \textit{TESS} Alert data from pipelines at the \textit{TESS} Science Office and at the \textit{TESS} Science Processing Operations Center.  This research has made use of the Exoplanet Follow-up Observation Program website, which is operated by the California Institute of Technology, under contract with the National Aeronautics and Space Administration under the Exoplanet Exploration Program.  Resources supporting this work were provided by the NASA High-End Computing (HEC) Program through the NASA Advanced Supercomputing (NAS) Division at Ames Research Center for the production of the SPOC data products.  This paper includes data collected by the \textit{TESS} mission, which are publicly available from the Mikulski Archive for Space Telescopes (MAST).
This paper is partially based on observations made with the Nordic Optical Telescope, operated by the Nordic Optical Telescope Scientific Association at the Observatorio del Roque de los Muchachos, La Palma, Spain, of the Instituto de Astrofisica de Canarias.

This research has made use of the NASA Exoplanet Archive, which is operated by the California Institute of Technology, under contract with the National Aeronautics and Space Administration under the Exoplanet Exploration Program.

%

\vspace{5mm}
\facilities{TESS, {\textsc{Minerva}}-Australis, NOT/FIES, LCOGT/NRES, Exoplanet Archive}


\software{AstroImageJ \citep{Collins:2017}, isochrones \citep{morton15}, ispec \citep{ispec2014,ispec2019}, Tapir \citep{Jensen:2013}, BANZAI \citep{McCully:2018}, Wotan \citep{wotan}, SPOC pipeline \citep{jenkins2016}, batman \citep{Kreidberg2015}, Exo-Striker \citep{Trifonov2019_es}  }








\clearpage

\bibliographystyle{aasjournal}





\end{document}